\documentclass[useAMS,usenatbib]{mn2e}
\usepackage{graphicx}
\usepackage{color}
\usepackage{ulem}
\def\apj{ApJ}
\def\apjs{ApJS}

\def\aap{A\&A}

\def\aj{AJ}
\def\mnras{MNRAS}

\def\pasj{PASJ}
\def\pasp{PASP}

\def\araa{Ann.Rev.Astron.Astrophys.}

\def\rmxaa{Revista Mexicana de Astronomia y Astrofisica}

\def\assl{Astrophysics and Space Science Library}
\def\assp{Astrophys. Space Sci. Proc.}
\def\aspc{Astronomical Societyof the Pacific}

\title[Monitoring broad emission in SBS 1420$+$540 and J1444$+$4840]
{Monitoring broad emission-line components in spectra of  the two low-metallicity dwarf compact star-forming galaxies SBS 1420$+$540 and J1444$+$4840}
\author[N. G. Guseva, T. X. Thuan, Y. I. Izotov]{N. G.\ Guseva$^{1}$,
T. X.\ Thuan$^{2}$ and Y. I.\ Izotov$^{1}$\\
                $^{1}$Bogolyubov Institute for Theoretical Physics,
                     Ukrainian National Academy of Sciences,
                     Metrologichna 14b, Kyiv 03143,  Ukraine,\\
                     nguseva@bitp.kiev.ua, yizotov@bitp.kiev.ua\\
                $^{2}$Astronomy Department, University of Virginia, 
                     P.O. Box 400325, Charlottesville, VA 22904-4325,\\
                     txt@virginia.edu\\
}
\begin{document}

\pagerange{\pageref{firstpage}--\pageref{lastpage}} \pubyear{2012}

\maketitle

\label{firstpage}

\begin{abstract}
We report the discovery of broad components with P-Cygni profiles of the hydrogen and
helium emission lines in the two low-redshift low-metallicity dwarf compact star-forming galaxies (SFG) SBS 1420$+$540 and J1444$+$4840. We found small stellar masses of 
10$^{6.24}$ and 10$^{6.59}$ M$_\odot$, low oxygen abundances 12 $+$ log O/H of 7.75 and 7.45,  high velocity dispersions reaching $\sigma$ $\sim$ 700  and $\sim$ 1200 km s$^{-1}$, high terminal velocities of the stellar wind of $\sim$1000 and $\sim$1000--1700 km s$^{-1}$, respectively, and large EW(H$\beta$) of $\sim$300 \AA\ for both. 
For SBS 1420$+$540, we succeeded in capturing an eruption phase by 
monitoring the variations of the broad-to-narrow component flux ratio. We observe a sharp increase of that ratio by a factor of 4 in 2017 and a decrease by about an order of magnitude in 2023. The peak luminosity of $\sim$ 10$^{40}$ erg s$^{-1}$ of the broad component in $L$(H$\alpha$) lasted for about 6 years out of a three-decades monitoring. This leads us to conclude that there is probably a Luminous Blue Variable candidate (LBVc) in this galaxy.
 As for J1444$+$4840, its very high $L$(H$\alpha$) of about 10$^{41}$ ergs s$^{-1}$, close to values observed in active galactic nuclei (AGNs) and Type IIn Supernovae (SNe), and the variability of no more than 20 per cent of the broad-to-narrow flux ratio  of the hydrogen and helium emission lines over a 8-year monitoring do not allow us to definitively conclude that it contains a LBVc.
 On the other hand, the possibility that the line variations are 
due to a long-lived stellar transient of type AGN/SN IIn cannot be ruled out.
\end{abstract}


\section{Introduction}\label{sec:INT}

    The Luminous Blue Variable (LBV) stage is a very short high luminous phase in the evolution of the most massive stars,
with masses greater than $\sim$20 -- 30M$_\odot$, during their thansition from early
young O stars to Wolf-Rayet (WR) stars and/or supernovae/black holes (SN/BHs)
\citep{Crowther2007,Solovyeva2020}.
      Besides being very hot and luminous massive stars, LBVs show significantly greater variability of brightness, observed both photometrically and spectroscopically, than other variable stars 
  \citep{HumphreysDavidson1994,Vink2012}.
They exhibit strong enhancement in emission and absorption lines, in the continuum level and in the blue
shape of the UV and optical continuum.
   The most prominent spectral features are broad components of hydrogen and often of 
helium lines, with P-Cygni profiles.
  The source of the broad emission 
can be stellar winds or eruptions of massive stars propagating in dense
circumstellar envelopes \citep*{HumphreysDavidson1994,Smith1994,Drissen1997,Drissen2001}.
  There are two types of variability in LBV stars:
   (1) the first type consists of moderate irregular quasi-periodic variations 
   in brightness,
due to variations in photospheric temperatures 
and to  strong and variable stellar wind mass-loss on timescales from several
years to several decades 
\citep[see e.g. ][]{Massey2000,Humphreys2013,Humphreys2017,Humphreys2019,Grassitelli2020,Weis2020}.
   Photometric variability of up to $\sim$0.5 -- 2 mag on a relatively short time-scale of several years is characteristic of the S Dor type of LBV stars.
     About a hundred such stars 
are known in the
Galaxy and in the nearest galaxies of the Local Group. Their proximity, especially for the LBVs in our Galaxy, allows a more detailed study of
the individual stars and their envelopes.  
    However, these young stars are located in the Galactic plane and are
subject to large extinction.
   In addition, the parameters of these massive stars, as well as those of the interstellar
environment in more distant galaxies, can be quite different. In particular,
the metallicity, the interstellar medium density, the star-formation rate (SFR) and the specific SFR (sSFR) can vary;
   (2) the second type of variability is found among a special class of rare LBVs with giant eruptions and brightening of more than 2.5 -- 3 mag,
 on timescales of up to thousands of years
\citep{DavidsonHumphreys1997,Smith2011,Vink2012,Weis2020}. 
   These are very rare events. Only two well established variable stars of this type are known in our Galaxy: $\eta$ Carinae and P-Cygni 
 \citep{Davidson1999,Lamers1983}.
   
     Until now, the origin of LBV variability remains uncertain and many questions concerning the nature of LBVs are left open: do all
 most massive stars  go through a LBV phase in their evolution?  
  Can a giant eruption and a S Dor-type variability occur in the same star
or only in different ones?  
We are also interested in investigating possible metallicity effects on the properties of LBV stars. 
The well-studied nearest star-forming dwarf galaxies Small Magellanic Cloud (SMC) and 
Large Magellanic Cloud (LMC) have respective metallicities of 0.2 and 0.5 that of the Sun, so they do not offer very metal-deficient environments. 
  Investigating the LBV phenomenon in star-forming galaxies (SFGs) with 
lower metallicities will allow us to better study the abundance dependence of 
stellar wind and mass loss properties of LBVs.

  As of now, these properties 
are still weakly constrained by existing observations 
 \citep{CrowtherHadfield2006,Bestenlehner2023,Vink2023}. That is because 
    only a few LBV/cLBV are now known in metal-poor distant SFGs
 \citep[see e.g.][and references therein]{Pustilnik2008,IzT09,Pustilnik2017,Annibali2019,Annibali2019a,Weis2020,Guseva2022}.

  The scarcity of known LBVs in distant galaxies is due to the difficulty of detecting them.
While it is impossible to obtain the spectrum of an individual LBV star in a distant galaxy, it is still quite a challenge to find signs of LBVs in the integral spectra of far-away galaxies. 
  The chance of finding a short-lived LBV in a distant galaxy increases with the number of massive 
stars in it. The latter depends in turn on the age of the starburst and hence on the  value of the
equivalent width of the H$\beta$ emission line.
  Up to thousands and tens of thousands young massive stars may be
present in the super-star clusters contained in SFGs with EW(H$\beta$)\,$>$\,100\AA, corresponding to a starburst age of less than 4 Myrs at a metallicity
12 + log(O/H) $<$ 8
\citep[see e.g. ][]{SchaererVacca1998,SchaererGus2000}.
    Broad component luminosities of 10$^{36}$ -- 10$^{38}$ ergs s$^{-1}$ are
observed in such SFGs which are attributed to massive stars with strong stellar
winds or to LBV/SN \citep{Drissen1997,Drissen2001,IzTG07,Guseva2022}.

   In this paper, we present long-slit spectrophotometric
observations with high signal-to-noise ratio of two compact dwarf SFGs, 
SBS 1420$+$544 and J1444$+$4840, which have respectively EW(H$\beta$) = 328 \AA\  and 288 \AA. 
SBS 1420$+$544 was selected from the Second Byurakan Survey (SBS) and J1444$+$4840 from the Sloan Digital Sky Survey (SDSS) Data Release 16 (DR16) \citep{Ahumada2020}. Both show 
strong emission lines with a strong broad component in the hydrogen and helium lines.
    These broad features are often used to find LBV stars in the integral spectra of galaxies in the local Universe. 

  The observations were obtained with the Large Binocular Telescope (LBT), equipped with the Multi-Object Double Spectrograph (MODS).
  In addition, SBS 1420$+$544 was also observed with the LBT Utility Camera in the Infrared (LUCI).
We use the LBT observations to derive  the physical and chemical properties of the two galaxies with high accuracy.
    In addition to the LBT data, we have also collected SDSS data and other observations of the two objects from previous studies. We have also added several new observations obtained with the 3.5m Apache Point Observatory (APO) telescope. This allows us to extend  our monitoring baseline of the two galaxies to as long a duration as possible.

  The LBT observations and data reduction of SBS\,1420$+$544 and J1444$+$4840 
are described in Section~\ref{sec:observations}.
  General properties of the studied galaxies are considered in subsection~\ref{sec:general}.
Subsection~\ref{sec:diagnostic} discusses their locations in various emission-line diagnostic diagrams.   
Derived element abundances are presented in subsection~\ref{subsec:abundances}.
  Subsection~\ref{sec:broad} describes the decomposition of the broad emission-line profiles of hydrogen and helium lines into various components. 
  In subsection~\ref{sec:signs} we discuss the evidence for the presence of LBV candidates in the two galaxies.
Subsubsections~\ref{sec:S1420} and ~\ref{sec:J1444} discuss the properties of SBS 1420$+$540 and J1444$+$4840 in detail.
    We summarize our main results in Sect.~\ref{sec:conclus}.

%%%%%%%%%%%%%%%%%%%%%%%%%%%%%%%%%%%%%%%%%%%%%%%%%%%%%%%%%%%
%  Table 1
%%%%%%%%%%%%%%%%%%%%%%%%%%%%%%%%%%%%%%%%%%%%%%%%%%%%%%%%%%%
\begin{table}
%\centering
\caption{Observed characteristics \label{tab1}}
\begin{tabular}{lrr} \hline
Parameter                 & SBS 1420$+$544&J1444$+$4840        \\ \hline
R.A.(J2000)               &  14:22:38.78& 14:44:59.01 \\%8433-57488-0714
Dec.(J2000)               & +54:14:09.76&+48:40:06.80\\
  $z$                     &  0.02113$\pm$0.00001&  0.06469$\pm$0.00005     \\
  $FUV$, mag              &   19.28$\pm$0.13&   22.01$\pm$0.41      \\
  $NUV$, mag              &   19.33$\pm$0.09&   22.18$\pm$0.42      \\
  $g$, mag                &   17.69$\pm$0.01&   20.71$\pm$0.03      \\
  $W1$, mag               &   16.42$\pm$0.06&  ...~~~~~~ \\
  $W2$, mag               &   15.06$\pm$0.06&   17.59$\pm$0.51\\
  $W3$, mag               &   10.42$\pm$0.06&   11.72$\pm$0.19\\
  $W4$, mag               &    7.35$\pm$0.09&  ...~~~~~~ \\
$D_L$, Mpc                &     92.5        &      292.5    \\
  $M_g$, mag$^\dag$        & $-$17.14$\pm$0.01&  $-$16.62$\pm$0.03    \\
log $M_\star$/M$_\odot$$^\ddag$&   6.24$\pm$0.33 &   6.59$\pm$0.35      \\
%log $M_\star$/M$_\odot$$^{\ddag\ddag}$&   6.29$\pm$0.06       \\
$L$(H$\beta$), erg s$^{-1}$$^*$&(1.2$\pm$0.1)$\times$10$^{40}$&(1.2$\pm$0.1)$\times$10$^{41}$\\
SFR, M$_\odot$yr$^{-1}$$^{\dag\dag}$  &     0.27$\pm$0.03&     2.65$\pm$0.03 \\
log sSFR, log yr$^{-1}$  &     $-$6.81&     $-$6.17 \\
\hline
  \end{tabular}

\noindent$^\dag$Corrected for Milky Way extinction.

\noindent$^\ddag$Derived from the extinction- and aperture-corrected SDSS 
spectrum.

%\noindent$^{\ddag\ddag}$Derived from the extinction- and aperture-corrected LBT 
%spectrum.

\noindent$^*$Corrected for extinction.%***Yura, draws attention to
%great disagreement between these data and those in Table 5.
%Read the explanation on page 9. At the same time, in Fig. 9b, the flux in SDSS is less than in LBT***

\noindent$^{\dag\dag}$Derived from the \citet{K98} relation using the extinction- 
and aperture-corrected H$\beta$ luminosity.
  \end{table}
%%%%%%%%%%%%%%%%%%%%%%%%%%%%%%%%%%%%%%%%%%%%%%%%%%%%%%%%%%%%%%%%%%%%

%%%%%%%%%%%%%%%%%%%%%%%%%%%%%%%%%%%%%%%%%%%%%%%%%%%%%%%%%%%
%  Table 2
%%%%%%%%%%%%%%%%%%%%%%%%%%%%%%%%%%%%%%%%%%%%%%%%%%%%%%%%%%%
\begin{table}
%\centering
\caption{Journal of LBT observations \label{tab2}}
\begin{tabular}{lcc} \hline
Parameter                 & SBS 1420$+$544&J1444$+$4840        \\ \hline
&\multicolumn{2}{c}{MODS} \\
Date           &2013-06-09&2019-04-02 \\
Exposure (s)   & 1800     & 7200 \\
Slit (arcsec)  & 60$\times$1.0      & 60$\times$1.2  \\
Airmass        & 1.29     & 1.04 \\
&\multicolumn{2}{c}{LUCI} \\
Date           &2016-06-16& ... \\
Exposure (s)   & 1440     & ... \\
Slit (arcsec)  & 60$\times$1.0      & ... \\
Airmass        & 1.10     & ... \\
\hline
  \end{tabular}
  \end{table}
%%%%%%%%%%%%%%%%%%%%%%%%%%%%%%%%%%%%%%%%%%%%%%%%%%%%%%%%%%%%%%%%%%%%

%%%%%%%%%%%%%%%%%%%%%%%%%%%%%%%%%%%%%%%%%%%%%%%%
%    Fig.1 diagnostic diagrams
%%%%%%%%%%%%%%%%%%%%%%%%%%%%%%%%%%%%%%%%%%%%%%%%
\begin{figure}
  \includegraphics[angle=-90,width=0.99\linewidth]{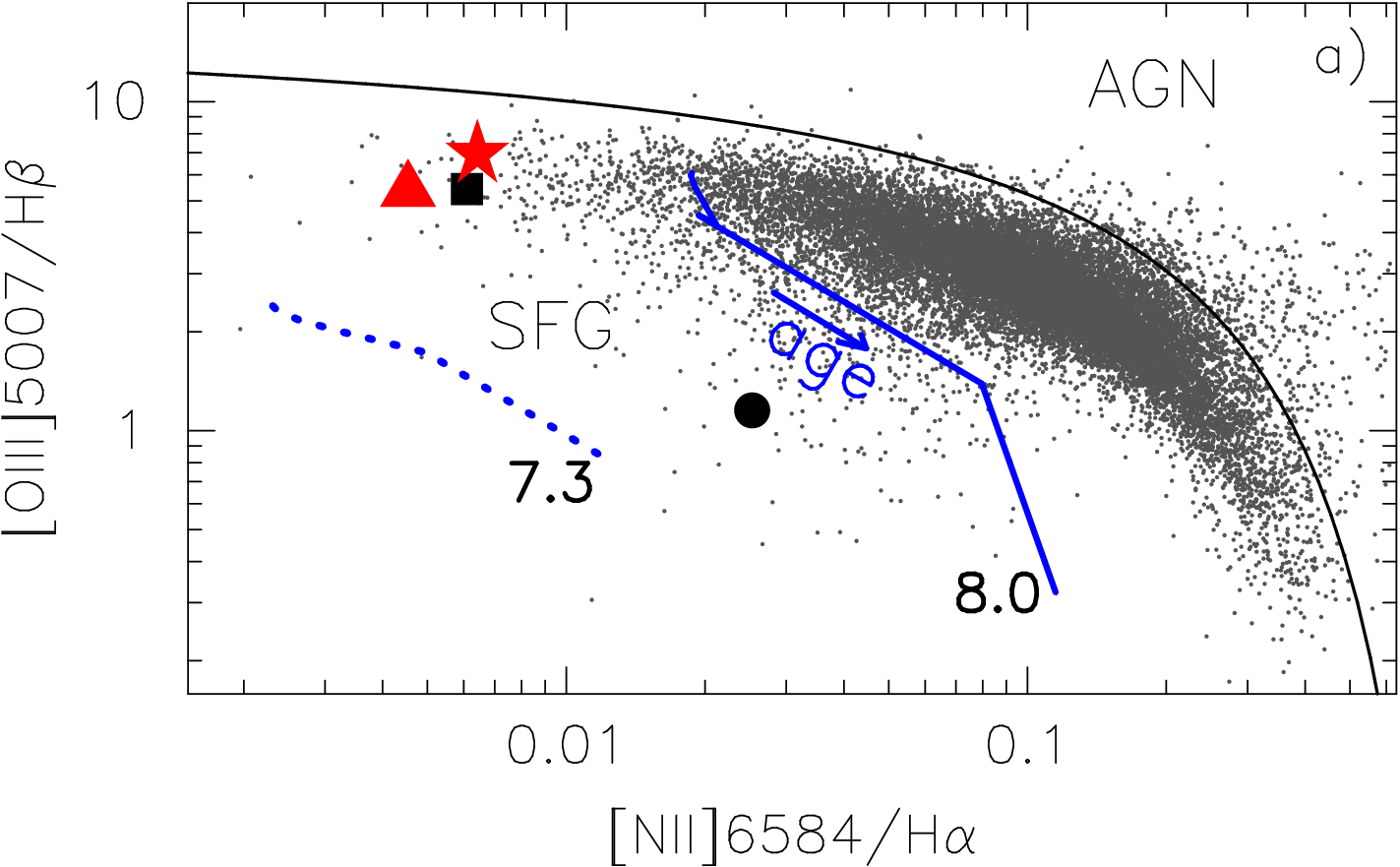}  
\includegraphics[angle=-90,width=0.99\linewidth]{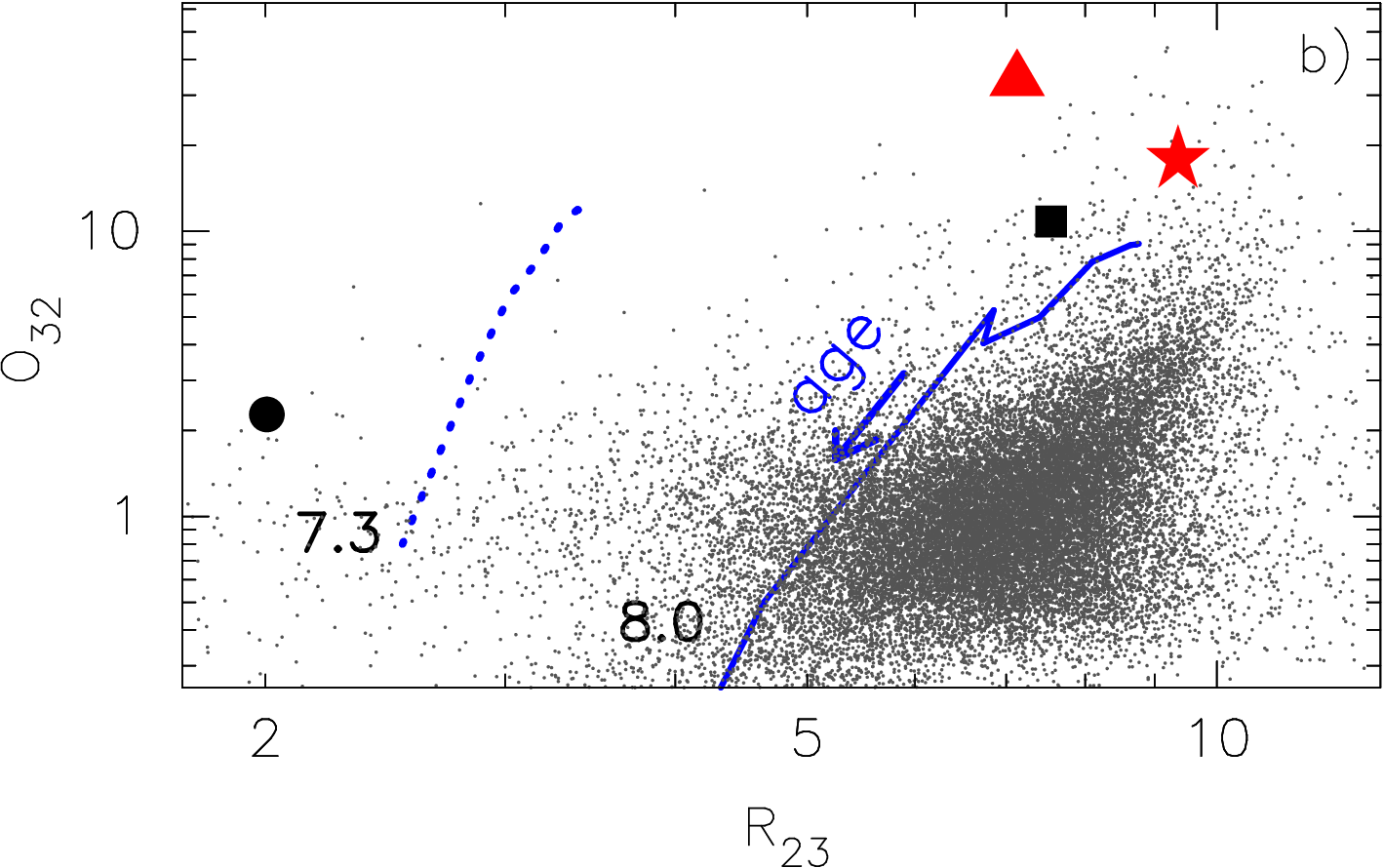}
\caption{{\bf a)} The Baldwin-Phillips-Terlevich (BPT) diagram \citep{BPT81} for
SFGs. {\bf b)} The O$_{32}$ -- R$_{23}$ diagram for SFGs where R$_{23}$ = 
([O~{\sc ii}]3727 + [O~{\sc iii}]4959 + [O~{\sc iii}]5007)/H$\beta$ and O32 = [O~{\sc iii}]5007/[O~{\sc ii}]3727. In both 
panels, SBS~1420$+$544 and J1444$+$4840 are shown by red filled star and red
filled triangle, respectively. PHL 293B and  H {\sc ii} region \#3 in DDO 68  with LBV candidates \citep{Pustilnik2008,IzT09,Guseva2022} 
are shown by black filled square and circle, respectively.
The compact SFGs from the SDSS \citep{I16c} are represented by grey dots.
 Blue dotted and blue thick solid 
lines show CLOUDY photoionized H~{\sc ii} region models for the relations obtained by \citet{Iz2018} using metallicities 
12 $+$ log O/H = 7.3 and 8.0, respectively. 
The directions of increasing starburst ages from 0 to 6 Myr are shown by arrows.
The black thin solid line in {\bf a)} separates SFGs from active galactic nuclei (AGN) \citep{K03}.
\label{fig1}}
\end{figure}
%%%%%%%%%%%%%%%%%%%%%%%%%%%%%%%%%%%%%%%%%%%%%%%%%

%%%%%%%%%%%%%%%%%%%%%%%%%%%%%%%%%%%%%%%%%%%%%%%%
%    Fig.2 LBT spectra of SBS 1420+544
%%%%%%%%%%%%%%%%%%%%%%%%%%%%%%%%%%%%%%%%%%%%%%%%
\begin{figure}
\includegraphics[angle=-90,width=0.99\linewidth]{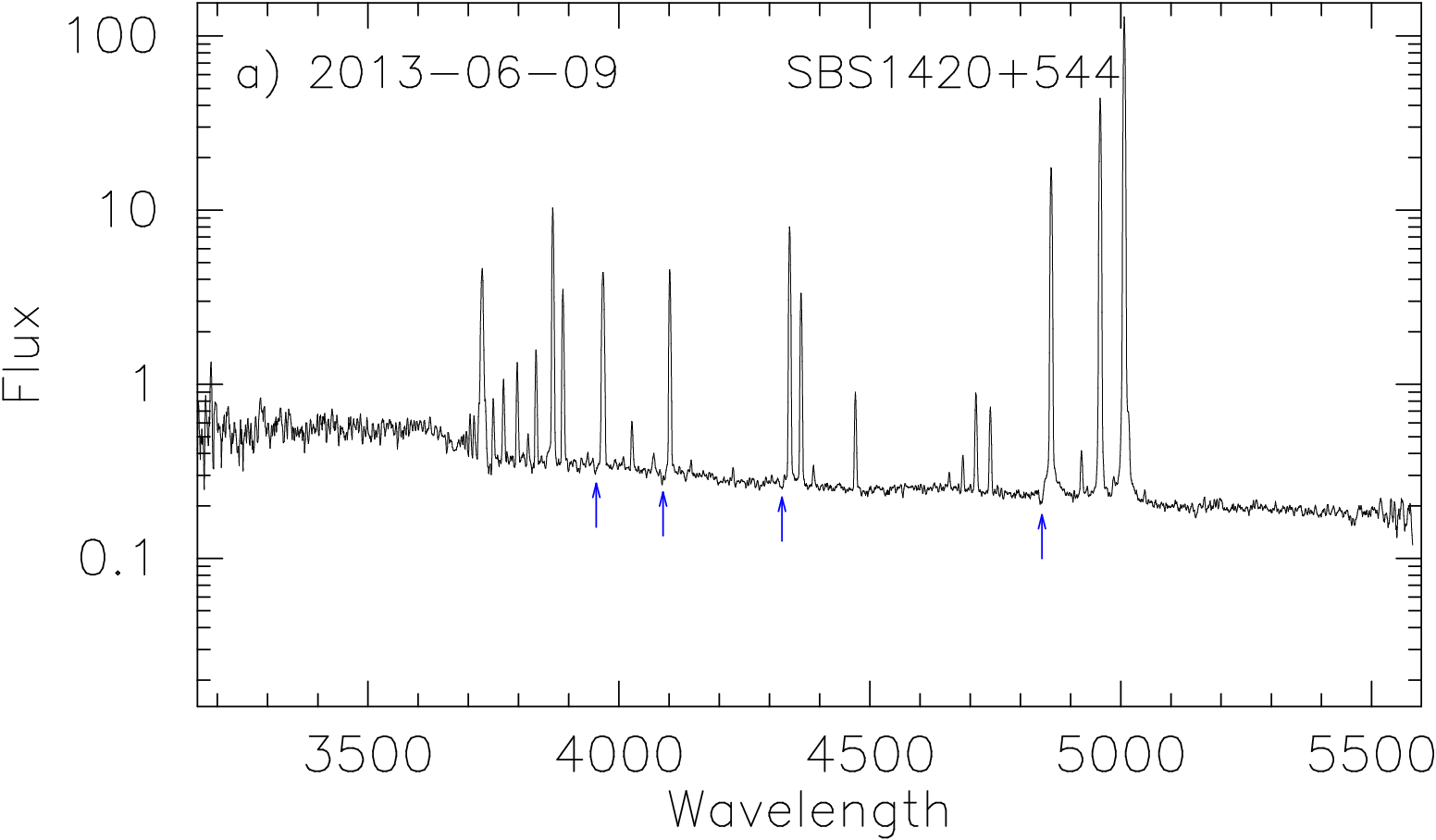}
\includegraphics[angle=-90,width=0.99\linewidth]{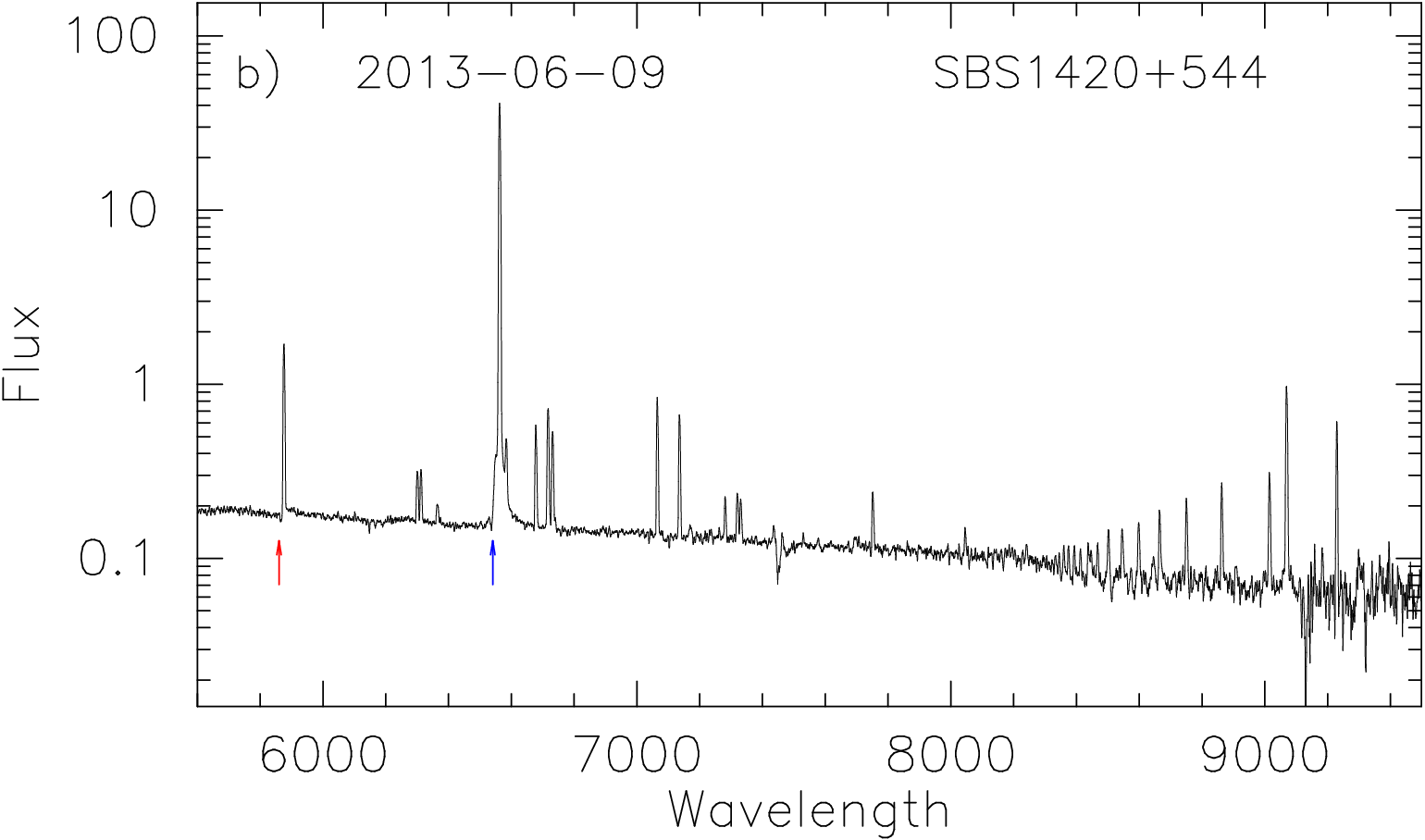}
\includegraphics[angle=-90,width=0.99\linewidth]{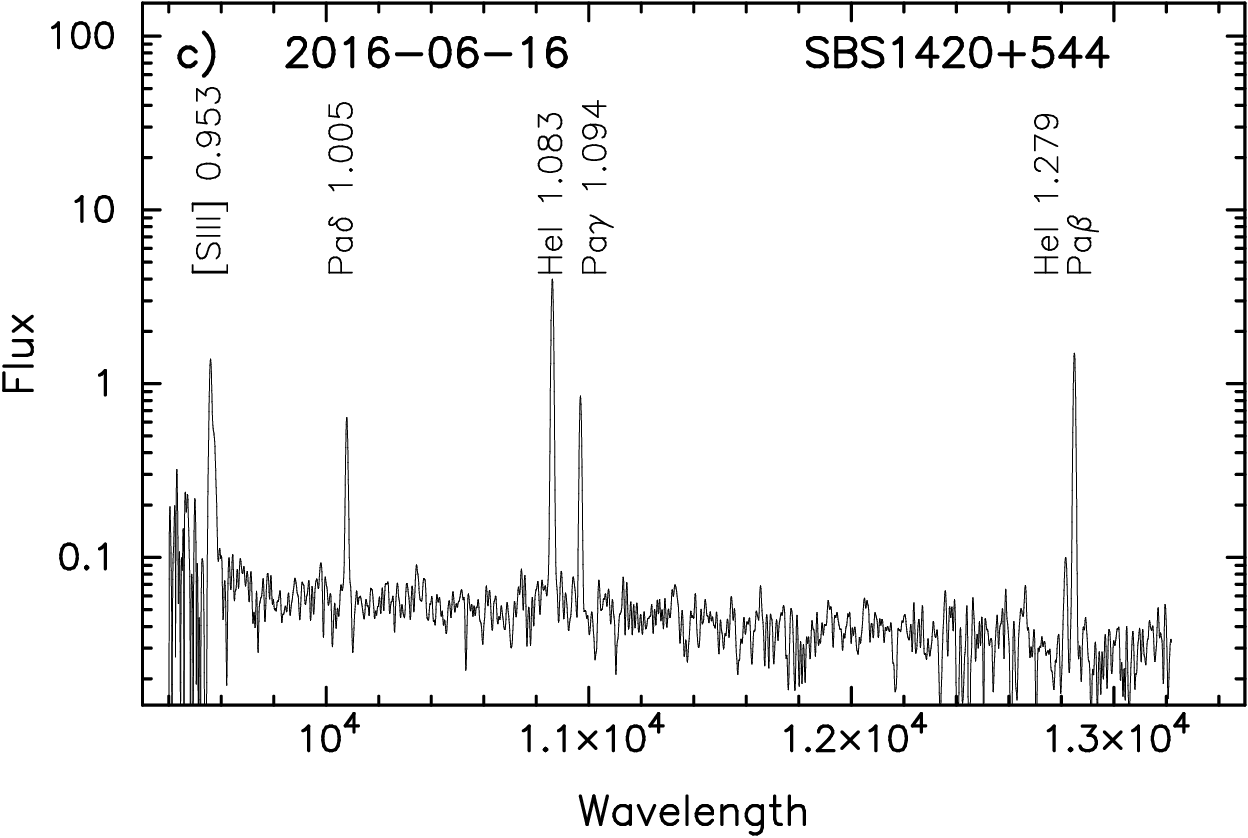}
\caption{The rest-frame {\bf a)} blue and {\bf b)} red LBT/MODS and {\bf c)}
LBT/LUCI spectra of SBS~1420$+$544 uncorrected for 
extinction. Blue absorption in P-Cygni profiles of hydrogen and helium 
lines are marked by blue and red arrows, respectively. Wavelengths are in \AA\ and fluxes are in 
units of 10$^{-16}$ erg s$^{-1}$ cm$^{-2}$ \AA$^{-1}$.}
\label{fig2}
\end{figure}
%%%%%%%%%%%%%%%%%%%%%%%%%%%%%%%%%%%%%%%%%%%%%%%%%

%%%%%%%%%%%%%%%%%%%%%%%%%%%%%%%%%%%%%%%%%%%%%%%%
%    Fig.3 LBT spectra of J1444+4840
%%%%%%%%%%%%%%%%%%%%%%%%%%%%%%%%%%%%%%%%%%%%%%%%
\begin{figure}
\includegraphics[angle=-90,width=0.99\linewidth]{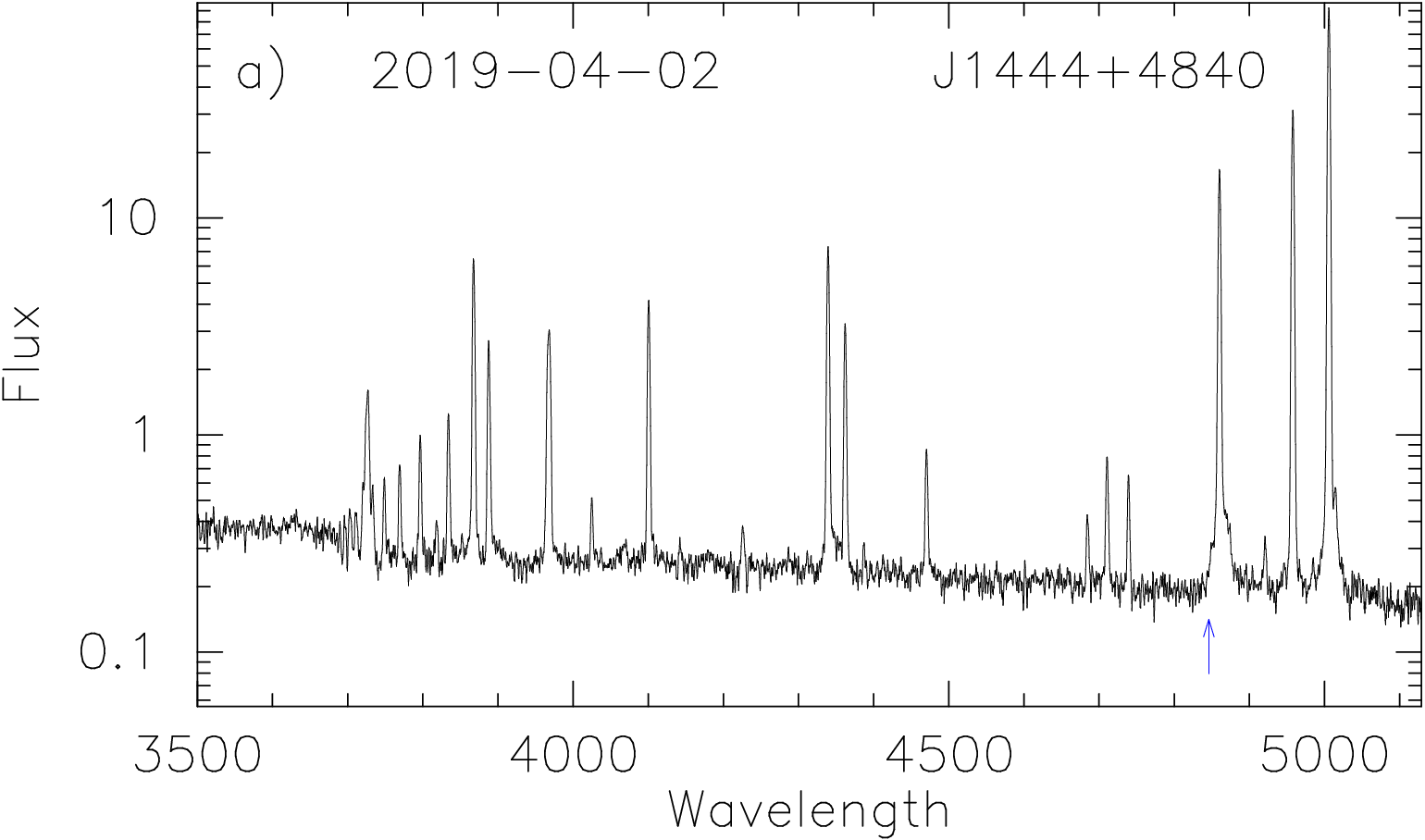}
\includegraphics[angle=-90,width=0.99\linewidth]{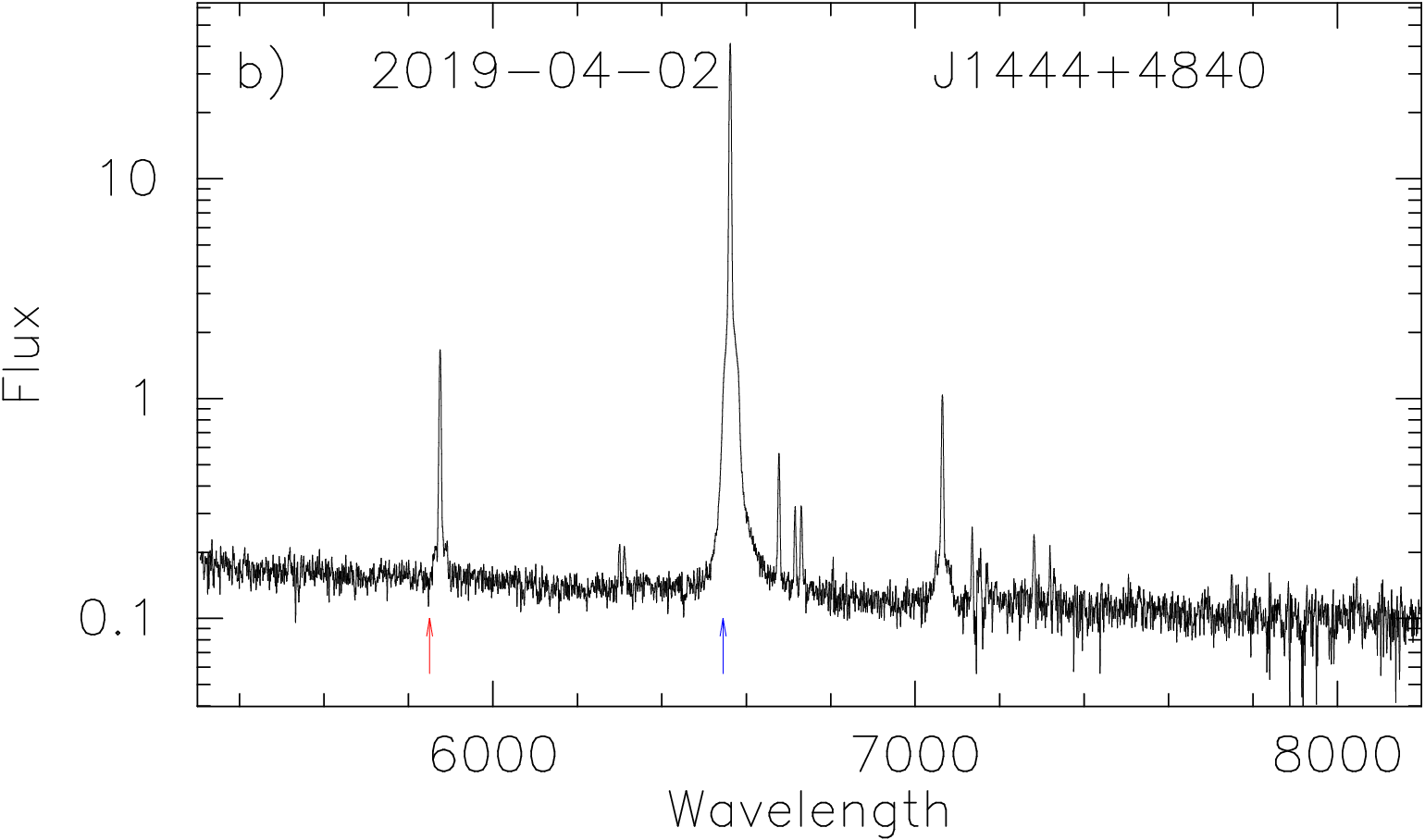}
\caption{The rest-frame LBT/MODS spectrum of J1444$+$4840 uncorrected for 
extinction. Blueward absorption in P-Cygni profiles of hydrogen
and helium lines is marked by blue and red arrows, respectively. Wavelengths are in \AA\ and fluxes are in 
units of 10$^{-16}$ erg s$^{-1}$ cm$^{-2}$ \AA$^{-1}$.}
\label{fig3}
\end{figure}
%%%%%%%%%%%%%%%%%%%%%%%%%%%%%%%%%%%%%%%%%%%%%%%%%

%%%%%%%%%%%%%%%%%%%%%%%%%%%%%%%%%%%%%%%%%%%%%%%%%%%%%%%%%%%
%  Table 3
%%%%%%%%%%%%%%%%%%%%%%%%%%%%%%%%%%%%%%%%%%%%%%%%%%%%%%%%%%%
\begin{table}
\caption{LBT extinction-corrected emission-line fluxes$^*$ \label{tab3}}
\begin{tabular}{lrr} \hline
 &\multicolumn{2}{c}{100$\times$$I$($\lambda$)/$I$(H$\beta$)}\\
Line&\multicolumn{1}{c}{SBS 1420$+$544}& \multicolumn{1}{c}{J1444$+$4840} \\ \hline
3187.74 He {\sc i}              &  5.85$\pm$3.22&  3.14$\pm$0.52\\
%3203.10 He {\sc ii}             &       ...     &  2.09$\pm$1.02\\
%3634.25 He {\sc i}              &  0.66$\pm$0.22&  0.76$\pm$0.25\\
%3686.83 H19                     &  0.54$\pm$0.13&  0.62$\pm$0.14\\
3691.55 H18                     &...~~~~~~&  0.79$\pm$0.14\\
3697.15 H17                     &  0.92$\pm$0.25&  1.04$\pm$0.17\\
3703.30 H16                     &  2.21$\pm$0.31&  1.66$\pm$0.18\\
3711.97 H15                     &  3.48$\pm$0.52&  1.99$\pm$0.34\\
3721.94 H14                     &  1.72$\pm$0.21&  2.45$\pm$0.20\\
3727.00 [O {\sc ii}]            & 38.87$\pm$1.28& 15.85$\pm$0.58\\
3734.37 H13                     &  2.37$\pm$0.23&  2.80$\pm$0.24\\
3750.15 H12                     &  4.15$\pm$0.44&  3.16$\pm$0.32\\
3770.63 H11                     &  5.89$\pm$0.44&  4.30$\pm$0.35\\
3797.90 H10                     &  7.27$\pm$0.43&  5.98$\pm$0.37\\
3819.64 He {\sc i}              &  1.06$\pm$0.18&  1.13$\pm$0.14\\
3835.39 H9                      &  8.81$\pm$0.44&  8.27$\pm$0.42\\
3868.76 [Ne {\sc iii}]          & 60.85$\pm$1.92& 47.57$\pm$1.54\\
3889.00 He {\sc i}+H8           & 20.49$\pm$0.72& 19.81$\pm$0.72\\
3968.00 [Ne {\sc iii}]+H7       & 36.65$\pm$1.18& 32.30$\pm$1.07\\
4009.26 He {\sc i}              &  0.35$\pm$0.10&...~~~~~~\\
4026.19 He {\sc i}              &  1.86$\pm$0.13&  2.11$\pm$0.17\\
4068.60 [S {\sc ii}]            &  0.72$\pm$0.12&...~~~~~~\\
4101.74 H$\delta$               & 27.03$\pm$0.86& 29.03$\pm$0.95\\
4120.84 He {\sc i}              &  0.22$\pm$0.07&...~~~~~~\\
4143.76 He {\sc i}              &  0.38$\pm$0.08&  0.68$\pm$0.10\\
4227.20 [Fe {\sc v}]            &  0.39$\pm$0.08&  1.29$\pm$0.11\\
4287.33 [Fe {\sc ii}]           &...~~~~~~&  0.62$\pm$0.12\\
4340.47 H$\gamma$               & 46.08$\pm$1.37& 49.53$\pm$1.51\\
4363.21 [O {\sc iii}]           & 18.43$\pm$0.56& 20.48$\pm$0.64\\
4387.93 He {\sc i}              &  0.56$\pm$0.08&  0.69$\pm$0.02\\
4471.48 He {\sc i}              &  3.73$\pm$0.14&  4.12$\pm$0.18\\
4562.50 [Mg {\sc i}]            &  0.15$\pm$0.06&...~~~~~~\\
4571.10 Mg {\sc i}]             &  0.20$\pm$0.07&...~~~~~~\\
4658.10 [Fe {\sc iii}]          &  0.32$\pm$0.05&...~~~~~~\\
4685.94 He {\sc ii}             &  0.85$\pm$0.07&  1.52$\pm$0.14\\
4712.00 [Ar {\sc iv}]+He {\sc i}&  3.92$\pm$0.14&  4.14$\pm$0.19\\
4740.20 [Ar {\sc iv}]           &  2.85$\pm$0.11&  3.04$\pm$0.16\\
4754.72 [Fe {\sc iii}]          &  0.18$\pm$0.04&...~~~~~~\\
4861.33 H$\beta$                &100.00$\pm$2.87&100.00$\pm$2.91\\
4921.93 He {\sc i}              &  1.11$\pm$0.07&  1.23$\pm$0.12\\
4958.92 [O {\sc iii}]           &234.57$\pm$6.71&181.06$\pm$5.23\\
4988.00 [Fe {\sc iii}]          &  0.52$\pm$0.05&...~~~~~~\\
5006.80 [O {\sc iii}]           &690.35$\pm$19.8&536.31$\pm$15.4\\
5015.68 He {\sc i}              &  0.91$\pm$0.04&  1.57$\pm$0.10\\
5199.00 [N {\sc i}]             &  0.19$\pm$0.05&...~~~~~~\\
5270.63 [Fe {\sc iii}]          &  0.13$\pm$0.05&...~~~~~~\\
5411.52 He {\sc ii}             &...~~~~~~&  0.35$\pm$0.08\\
5517.71 [Cl {\sc iii}]          &  0.34$\pm$0.12&  0.23$\pm$0.05\\
5537.88 [Cl {\sc iii}]          &  0.17$\pm$0.06&...~~~~~~\\
5875.60 He {\sc i}              & 10.86$\pm$0.33& 11.46$\pm$0.36\\
6300.30 [O {\sc i}]             &  1.15$\pm$0.05&  0.46$\pm$0.04\\
6312.10 [S {\sc iii}]           &  1.13$\pm$0.05&  0.54$\pm$0.04\\
6363.80 [O {\sc i}]             &  0.44$\pm$0.04&  0.19$\pm$0.01\\
6548.10 [N {\sc ii}]            &  0.95$\pm$0.04&...~~~~~~\\
6562.80 H$\alpha$               &278.81$\pm$8.63&275.59$\pm$8.64\\
6583.40 [N {\sc ii}]            &  1.79$\pm$0.07&  1.25$\pm$0.07\\
\hline
  \end{tabular}
  \end{table}

\begin{table}
\contcaption{LBT extinction-corrected emission-line fluxes \label{tab2a}}
\begin{tabular}{lrr} \hline
 &\multicolumn{2}{c}{100$\times$$I$($\lambda$)/$I$(H$\beta$)}\\
Line&\multicolumn{1}{c}{SBS 1420$+$544}& \multicolumn{1}{c}{J1444$+$4840} \\ \hline
6678.10 He {\sc i}              &  2.92$\pm$0.10&  2.87$\pm$0.11\\
6716.40 [S {\sc ii}]            &  4.17$\pm$0.14&  1.17$\pm$0.06\\
6730.80 [S {\sc ii}]            &  2.77$\pm$0.10&  1.29$\pm$0.06\\
7065.30 He {\sc i}              &  4.33$\pm$0.14&  5.63$\pm$0.19\\
7135.80 [Ar {\sc iii}]          &  3.49$\pm$0.12&  1.58$\pm$0.06\\
7170.62 [Ar {\sc iv}]           &  0.19$\pm$0.03&...~~~~~~\\
7237.00 [Ar {\sc iv}]           &  0.12$\pm$0.03&...~~~~~~\\
7262.76 [Ar {\sc iv}]           &  0.10$\pm$0.03&...~~~~~~\\
7281.35 He {\sc i}              &  0.60$\pm$0.04&  0.78$\pm$0.05\\
7319.90 [O {\sc ii}]            &  0.81$\pm$0.05&  0.52$\pm$0.05\\
7330.20 [O {\sc ii}]            &  0.67$\pm$0.04&  0.38$\pm$0.03\\
7751.12 [Ar {\sc iii}]          &  0.85$\pm$0.05&  0.26$\pm$0.03\\
8045.63 [Cl {\sc iv}]           &  0.27$\pm$0.04&...~~~~~~\\
8345.44 P23                     &  0.14$\pm$0.03&...~~~~~~\\
8359.01 P22                     &  0.20$\pm$0.04&...~~~~~~\\
8374.48 P21                     &  0.24$\pm$0.04&...~~~~~~\\
8392.40 P20                     &  0.24$\pm$0.04&...~~~~~~\\
8413.32 P19                     &  0.19$\pm$0.04&...~~~~~~\\
8437.96 P18                     &  0.28$\pm$0.04&...~~~~~~\\
8446.34 O {\sc i}               &  0.21$\pm$0.04&...~~~~~~\\
8467.26 P17                     &  0.26$\pm$0.04&  0.37$\pm$0.05\\
8502.49 P16                     &  0.55$\pm$0.05&  0.54$\pm$0.06\\
8545.38 P15                     &  0.56$\pm$0.06&  0.44$\pm$0.06\\
8598.39 P14                     &  0.68$\pm$0.06&  0.52$\pm$0.05\\
8665.02 P13                     &  0.77$\pm$0.06&  0.72$\pm$0.06\\
8750.47 P12                     &  0.96$\pm$0.06&  0.99$\pm$0.07\\
8862.79 P11                     &  1.13$\pm$0.07&  1.47$\pm$0.11\\
9014.91 P10                     &  1.54$\pm$0.09&  2.25$\pm$0.16\\
9069.00 [S {\sc iii}]           &  5.74$\pm$0.23&  2.46$\pm$0.16\\
9229.02 P9                      &  2.92$\pm$0.18&  1.79$\pm$0.15\\
9530.60 [S {\sc iii}]           & 15.16$\pm$0.60&...~~~~~~\\
9545.98 P8                      &  2.37$\pm$0.18&...~~~~~~\\
10050.00 P$\delta$              &  6.09$\pm$0.25&...~~~~~~\\
10829.00 He~{\sc i}             & 39.97$\pm$1.57&...~~~~~~\\
10941.00 P$\gamma$              &  7.74$\pm$0.32&...~~~~~~\\
12790.00 He~{\sc i}             &  0.86$\pm$0.06&...~~~~~~\\
12821.00 P$\beta$               & 14.20$\pm$0.59&...~~~~~~\\
 \\
 $C$(H$\beta$)$^{\rm a}$         &\multicolumn{1}{c}{0.205$\pm$0.037}&\multicolumn{1}{c}{0.275$\pm$0.037}\\
$F$(H$\beta$)$^{\rm b}$         &\multicolumn{1}{c}{76.07$\pm$2.20}&\multicolumn{1}{c}{59.94$\pm$2.20}\\
EW(H$\beta$)$^{\rm c}$          &\multicolumn{1}{c}{328.0$\pm$1.0}&\multicolumn{1}{c}{288.0$\pm$1.0}\\
EW(abs)$^{\rm c}$               &\multicolumn{1}{c}{2.4$\pm$0.5}&\multicolumn{1}{c}{0.8$\pm$0.4}\\
\hline
  \end{tabular}

\hbox{* Note that for bright hydrogen lines only nebular narrow}

\hbox{emission components were measured.}

\hbox{$^{\rm a}$Extinction coefficient, derived from the observed nebular} 

\hbox{hydrogen Balmer lines.}

\hbox{$^{\rm b}$Observed flux of narrow H$\beta$ in units of 10$^{-16}$ erg s$^{-1}$ cm$^{-2}$. }

\hbox{$^{\rm c}$Rest-frame equivalent width in \AA.}

  \end{table}
%%%%%%%%%%%%%%%%%%%%%%%%%%%%%%%%%%%%%%%%%%%%%%%%%%%%%%%%%%%%%%%%%%%%

%%%%%%%%%%%%%%%%%%%%%%%%%%%%%%%%%%%%%%%%%%%%%%%%%%%%%%%%%%%
%  Table 4
%%%%%%%%%%%%%%%%%%%%%%%%%%%%%%%%%%%%%%%%%%%%%%%%%%%%%%%%%%%
\begin{table}
  \caption{Electron temperatures, electron number densities and element
    abundances from LBT observations \label{tab4}}
\begin{tabular}{lccccc} \hline
Property                             &SBS 1420$+$544 &J1444$+$4840 \\ \hline
$T_{\rm e}$(O {\sc iii}), K          &17500$\pm$350 &21600$\pm$550       \\
$T_{\rm e}$(O {\sc ii}), K           &15100$\pm$280 &15600$\pm$360       \\
$T_{\rm e}$(S {\sc iii}), K          &16200$\pm$290 &19300$\pm$450       \\
$N_{\rm e}$(S {\sc ii}), cm$^{-3}$    &10$\pm$10     &970$\pm$240         \\ \\
O$^+$/H$^+$$\times$10$^6$            &3.438$\pm$0.200&1.408$\pm$0.096 \\
O$^{2+}$/H$^+$$\times$10$^5$          &5.273$\pm$0.279&2.640$\pm$0.159 \\
O$^{3+}$/H$^+$$\times$10$^6$          &0.499$\pm$0.051&0.442$\pm$0.050 \\
O/H$\times$10$^5$                   &5.667$\pm$0.280&2.825$\pm$0.159 \\
12+log(O/H)                         &7.753$\pm$0.021&7.451$\pm$0.024     \\ \\
N$^+$/H$^+$$\times$10$^7$            &1.298$\pm$0.058&0.860$\pm$0.050 \\
ICF(N)                              &14.465 &17.779 \\
N/H$\times$10$^6$                   &1.878$\pm$0.095 &1.529$\pm$0.101 \\
log(N/O)                            &$-$1.480$\pm$0.031~~~&$-$1.267$\pm$0.038~~~\\ \\
Ne$^{2+}$/H$^+$$\times$10$^5$        &1.065$\pm$0.060 &0.508$\pm$0.030 \\
ICF(Ne)                             &1.019 &1.027 \\
Ne/H$\times$10$^5$                  &1.085$\pm$0.064 &0.522$\pm$0.032 \\
log(Ne/O)                           &$-$0.718$\pm$0.034~~~&$-$0.733$\pm$0.036~~~\\ \\
S$^+$/H$^+$$\times$10$^8$            &6.580$\pm$0.238 &2.440$\pm$0.129 \\
S$^{2+}$/H$^+$$\times$10$^7$         &4.628$\pm$0.293 &1.380$\pm$0.129 \\
ICF(S)                              &2.060 &2.895 \\
S/H$\times$10$^6$                   &1.089$\pm$0.061 &0.470$\pm$0.038 \\
log(S/O)                            &$-$1.716$\pm$0.032~~~&$-$1.779$\pm$0.043~~~\\ \\
Cl$^{2+}$/H$^+$$\times$10$^8$        &1.114$\pm$0.292 &0.628$\pm$0.111 \\
ICF(Cl)                             &1.529 &1.507 \\
Cl/H$\times$10$^8$                  &1.702$\pm$0.447 &0.946$\pm$0.168 \\
log(Cl/O)                           &$-$3.522$\pm$0.116~~~&$-$3.475$\pm$0.081~~~\\ \\
Ar$^{2+}$/H$^+$$\times$10$^7$        &1.229$\pm$0.050 &0.430$\pm$0.018 \\
Ar$^{3+}$/H$^+$$\times$10$^7$        &1.969$\pm$0.102 &1.324$\pm$0.083 \\
ICF(Ar)                             &1.662 &1.872 \\
Ar/H$\times$10$^7$                  &2.042$\pm$0.189 &0.805$\pm$0.159 \\
log(Ar/O)                           &$-$2.443$\pm$0.046~~~&$-$2.546$\pm$0.089~~~\\ \\
Fe$^{2+}$/H$^+$$\times$10$^7$        &0.589$\pm$0.094 & ... \\
ICF(Fe)                             &21.789 & ... \\
Fe/H$\times$10$^6$                  &1.284$\pm$0.206 & ... \\
log(Fe/O)                           &$-$1.645$\pm$0.073~~~& ... \\
\hline
  \end{tabular}
  \end{table}
%%%%%%%%%%%%%%%%%%%%%%%%%%%%%%%%%%%%%%%%%%%%%%%%%%%%%%%%%%%%%%%%%%%%

%%%%%%%%%%%%%%%%%%%%%%%%%%%%%%%%%%%%%%%%%%%%%%%%
%    Fig.4 LBT, Decomposition of Hb, Ha (SBS 1420+544)
%%%%%%%%%%%%%%%%%%%%%%%%%%%%%%%%%%%%%%%%%%%%%%%%
\begin{figure}
\includegraphics[angle=-90,width=0.99\linewidth]{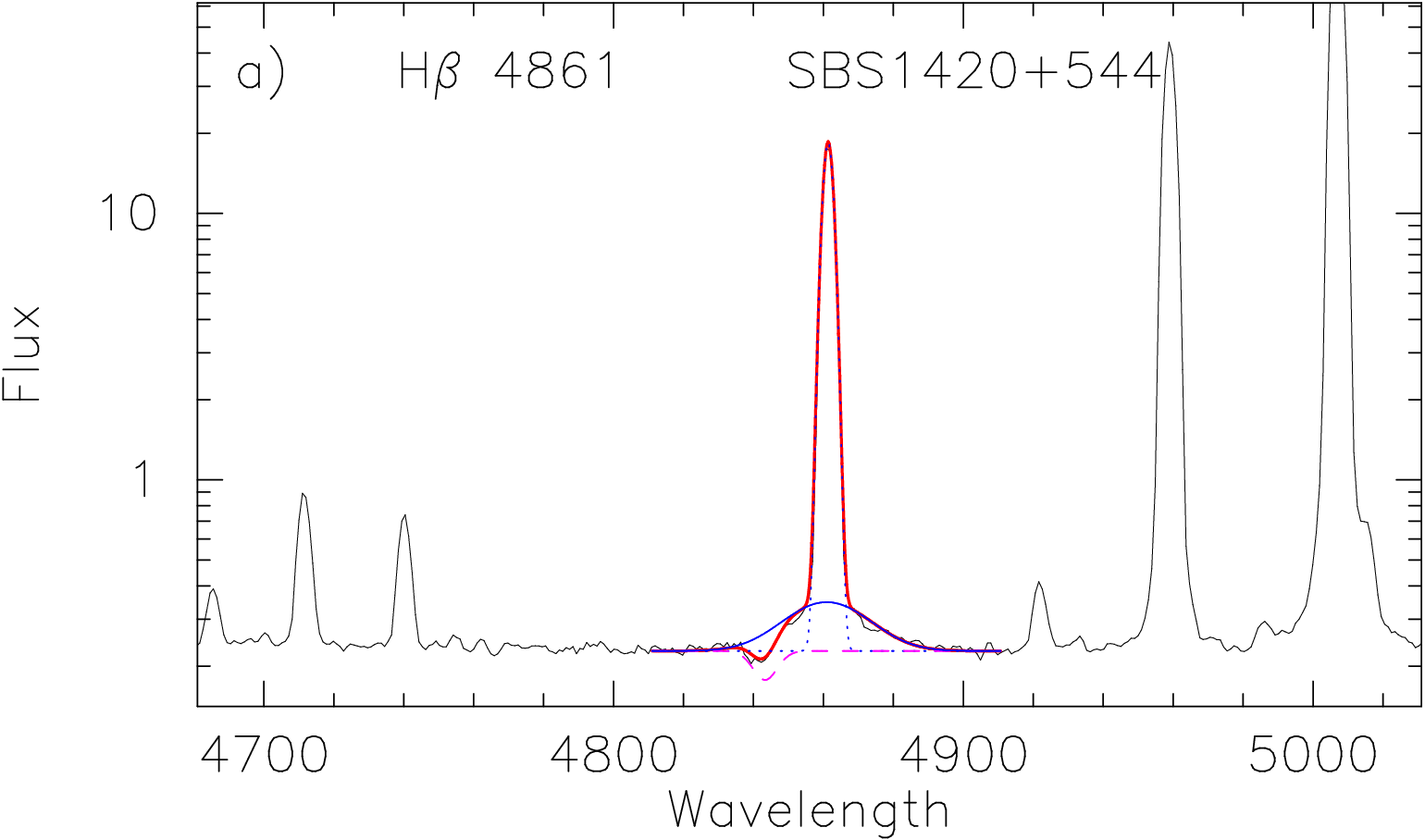}
\includegraphics[angle=-90,width=0.99\linewidth]{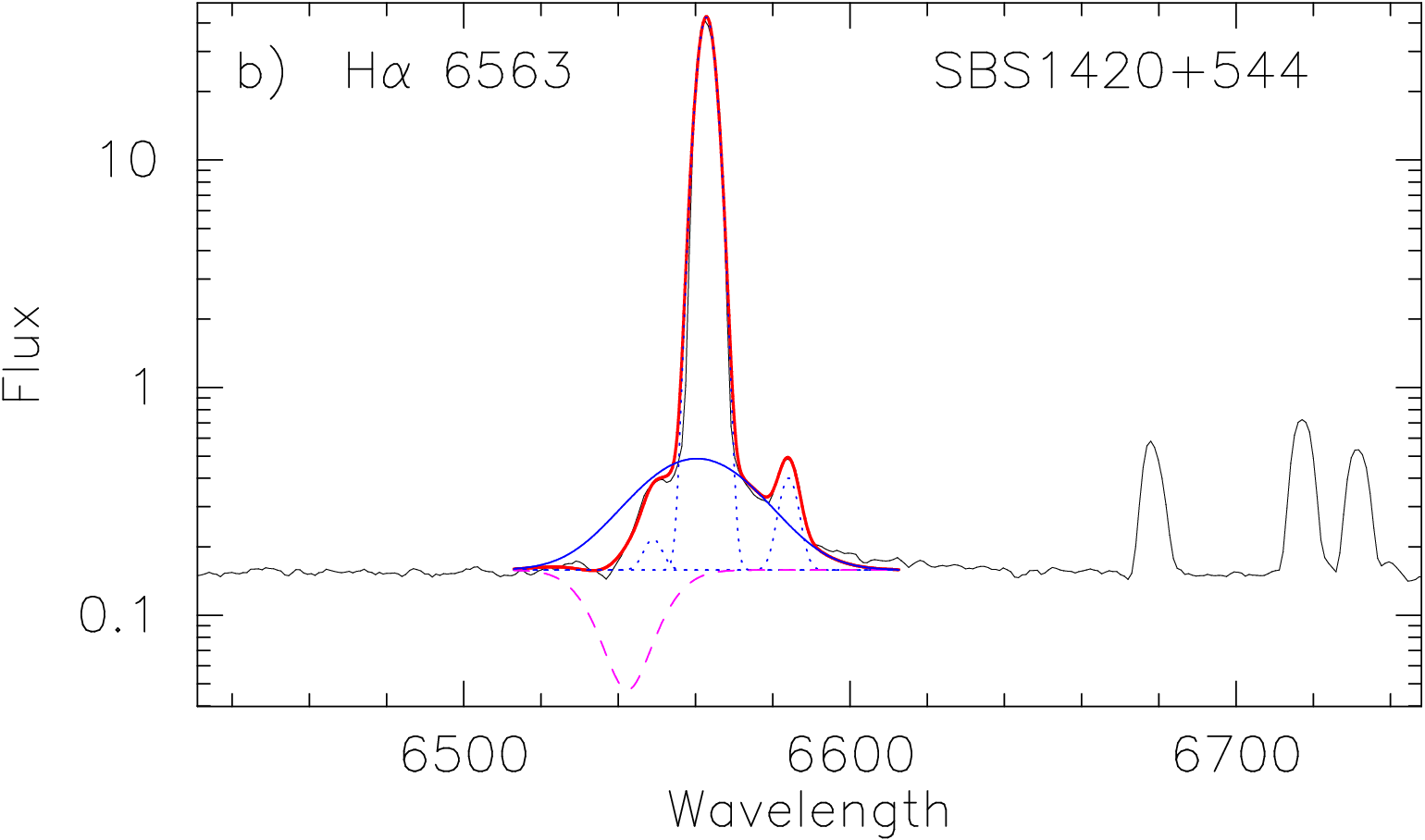}
\caption{Decomposition in the LBT spectrum by gaussians of {\bf a)} H$\beta$ and {\bf b)} H$\alpha$  narrow and 
broad emission lines and absorption lines of SBS~1420$+$544 (blue dotted and solid lines for emission, and magenta dashed lines for absorption, respectively). Additionally, two blue dotted lines in {\bf b)} also represent [N {\sc ii}]$\lambda$6548 and $\lambda$6583\AA\ emission lines.
The fitted profiles in {\bf a)} and {\bf b)} are shown by the red solid lines whereas the observed
spectra are represented by black solid lines. Wavelengths are in \AA\ and fluxes 
are in units of 10$^{-16}$ erg s$^{-1}$ cm$^{-2}$ \AA$^{-1}$.}
\label{fig4}
\end{figure}
%%%%%%%%%%%%%%%%%%%%%%%%%%%%%%%%%%%%%%%%%%%%%%%%%

%%%%%%%%%%%%%%%%%%%%%%%%%%%%%%%%%%%%%%%%%%%%%%%%
%    Fig.5 LBT, Decomposition Hβ,Hα,HeIλ5876,HeIλ7065 (J1444+4840)
%%%%%%%%%%%%%%%%%%%%%%%%%%%%%%%%%%%%%%%%%%%%%%%%
\begin{figure}
\includegraphics[angle=-90,width=0.99\linewidth]{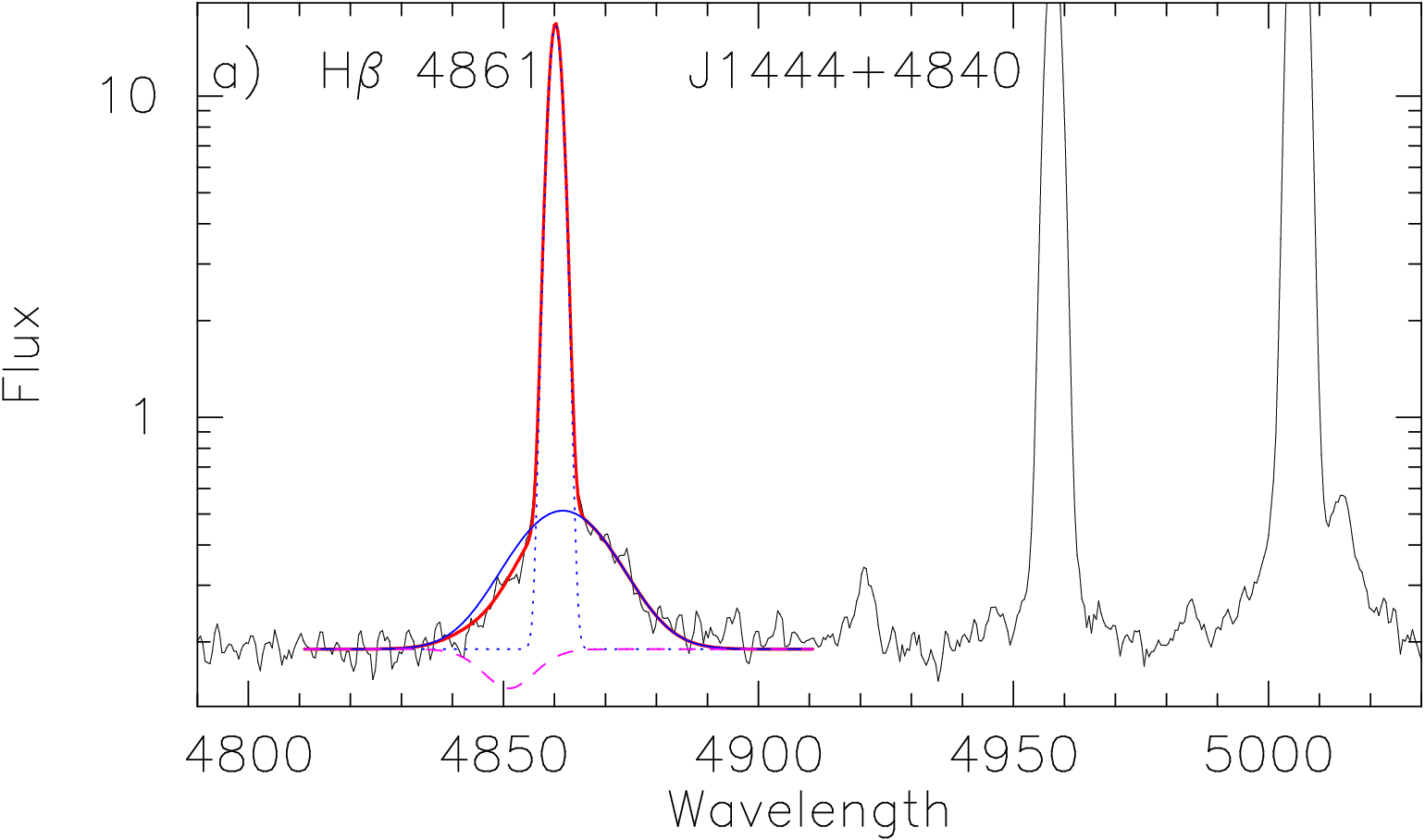}  
\includegraphics[angle=-90,width=0.99\linewidth]{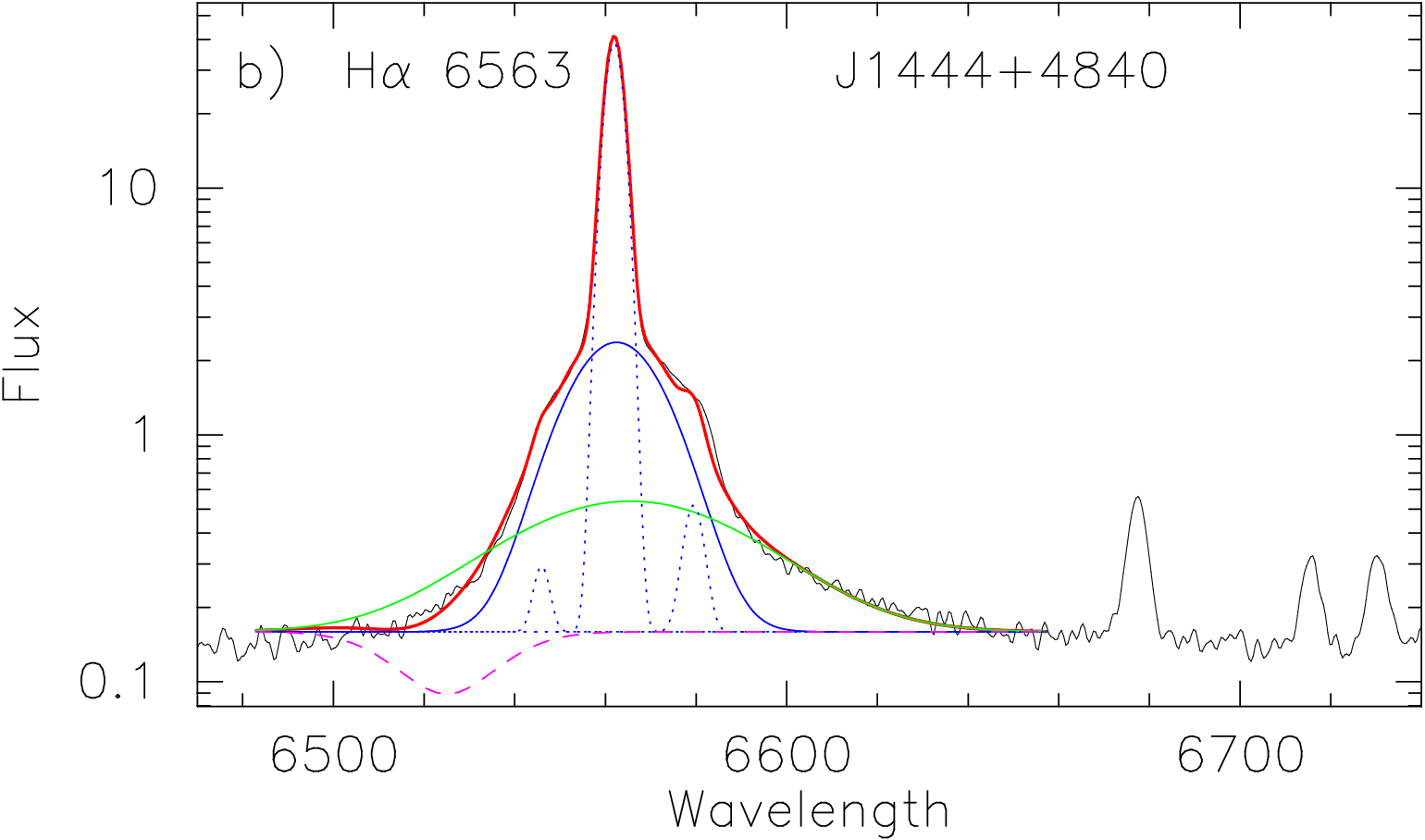}
\includegraphics[angle=-90,width=0.99\linewidth]{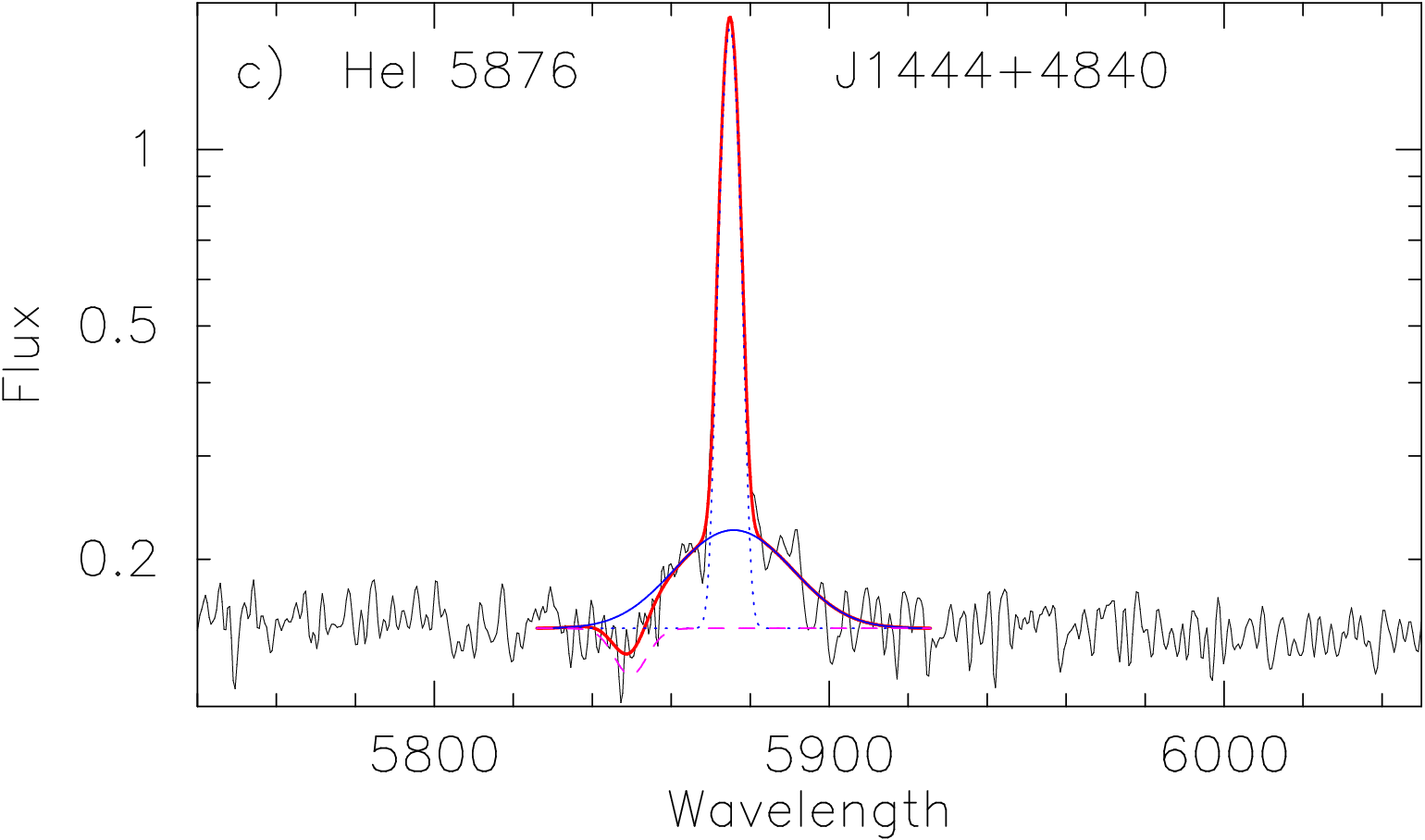}
\includegraphics[angle=-90,width=0.99\linewidth]{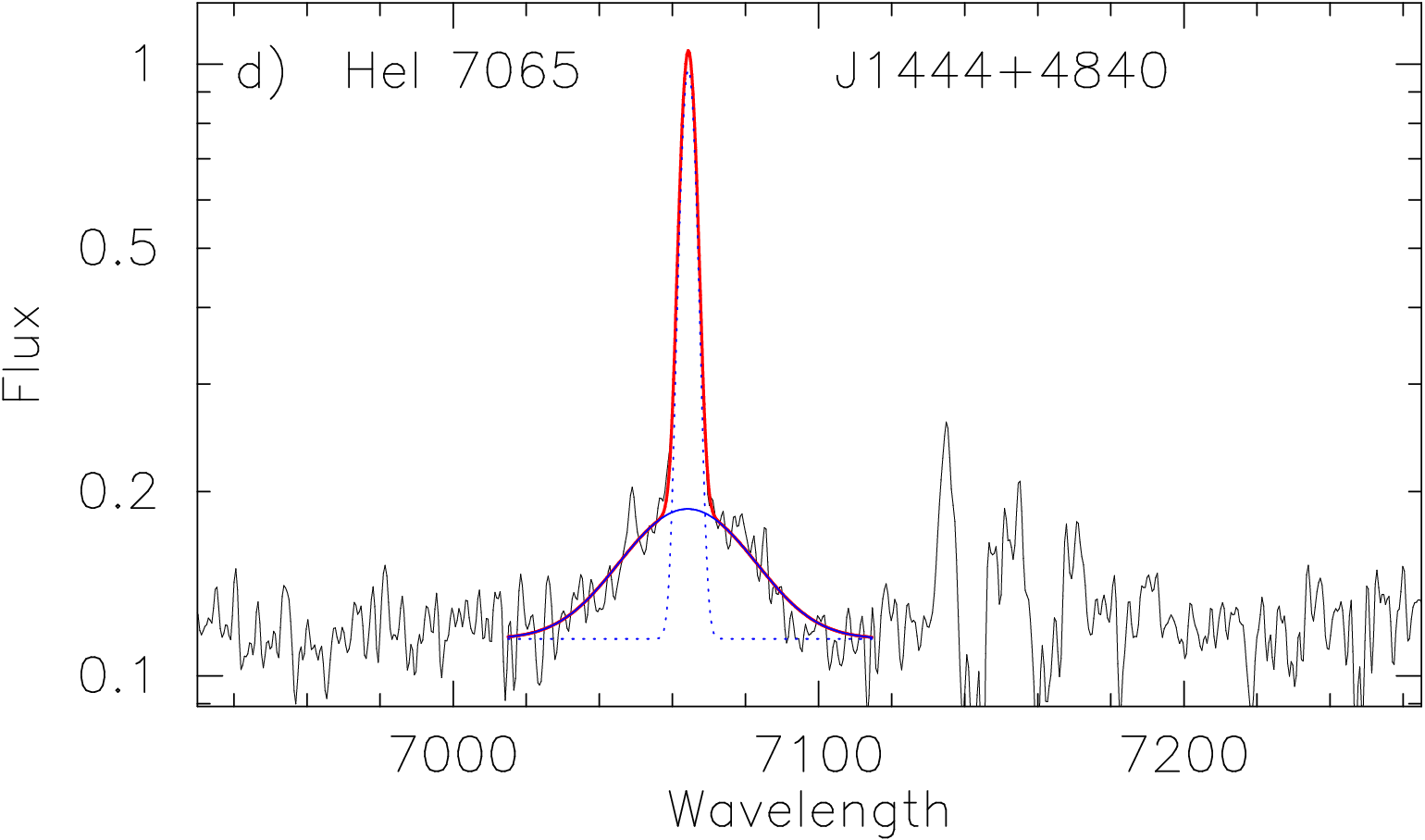}
\caption{Decomposition of {\bf a)} H$\beta$, {\bf b)} H$\alpha$,
{\bf c)} He~{\sc i} $\lambda$5876 and {\bf d)} He~{\sc i} $\lambda$7065 emission lines by gaussians of the LBT spectrum of J1444$+$4840.
Very broad component in the case of H$\alpha$ is shown by green solid line.
The rest of the designations in the figure are the same as in Fig.~\ref{fig4}.
}
\label{fig5}
\end{figure}
%%%%%%%%%%%%%%%%%%%%%%%%%%%%%%%%%%%%%%%%%%%%%%%%%

%%%%%%%%%%%%%%%%%%%%%%%%%%%%%%%%%%%%%%%%%%%%%%%%
%    Fig.6 Ha 1420 spectra
%%%%%%%%%%%%%%%%%%%%%%%%%%%%%%%%%%%%%%%%%%%%%%%%
\begin{figure*}
   \includegraphics[angle=-90,width=0.8\linewidth]{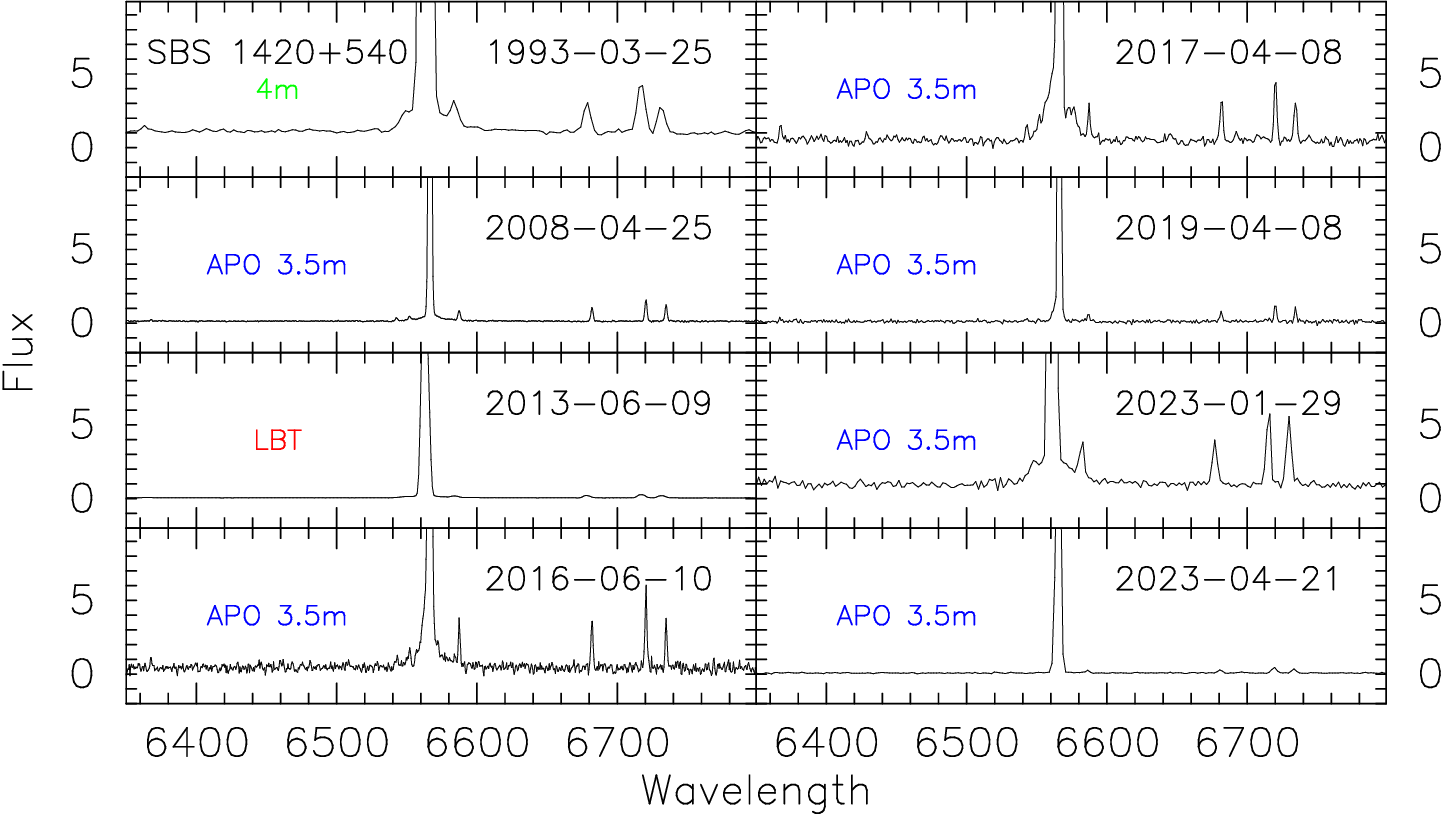} 
\caption{The rest-frame H$\alpha$ emission line profiles in the SBS 1420$+$544
  spectra at different epochs from collection of our observations carried out with 4m KPNO, LBT/MODS and 3.5m APO telescopes. 
 Wavelengths are
in \AA\ and fluxes are in units of 10$^{-16}$ erg s$^{-1}$ cm$^{-2}$ \AA$^{-1}$.}
\label{fig6}
\end{figure*}
%%%%%%%%%%%%%%%%%%%%%%%%%%%%%%%%%%%%%%%%%%%%%%%%%

%%%%%%%%%%%%%%%%%%%%%%%%%%%%%%%%%%%%%%%%%%%%%%%%
%    Fig.7 Ha 1444 spectra
%%%%%%%%%%%%%%%%%%%%%%%%%%%%%%%%%%%%%%%%%%%%%%%%
\begin{figure*}
   \includegraphics[angle=-90,width=0.8\linewidth]{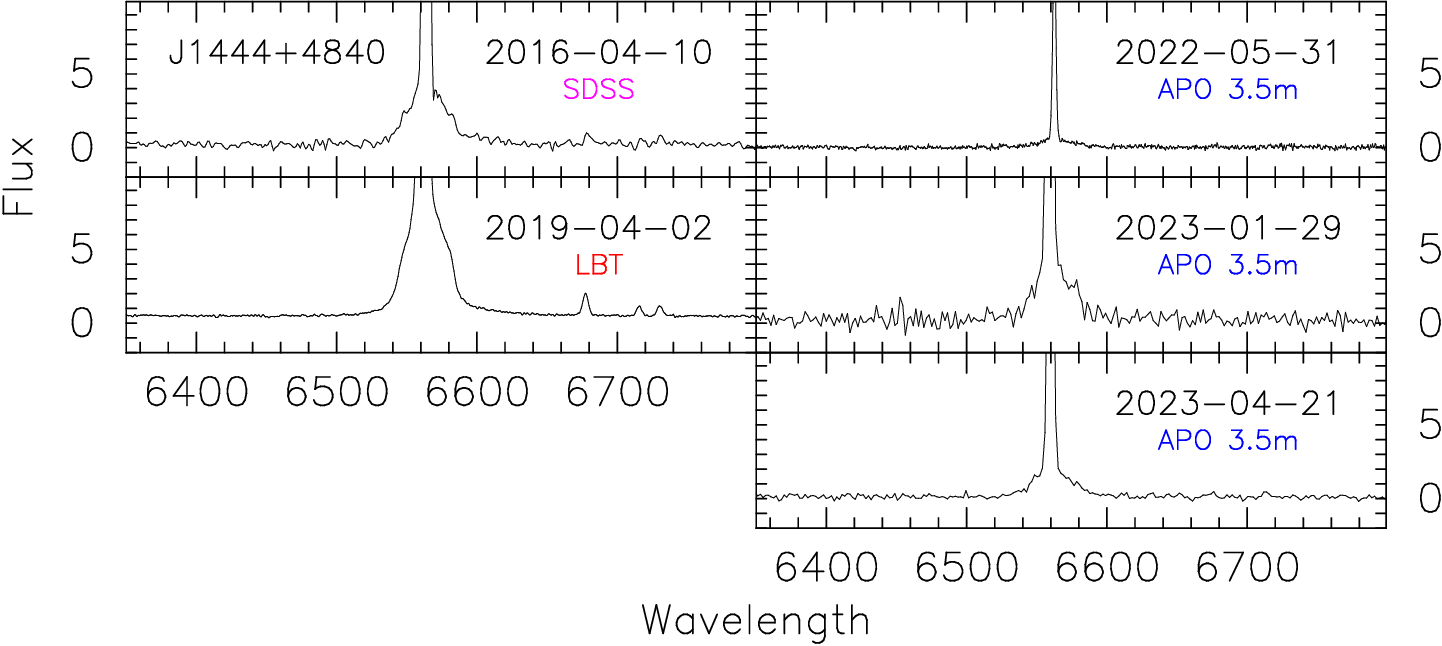} 
\caption{The rest-frame H$\alpha$ emission line profiles in the J1444$+$4840
  spectra at the different epochs, observed with the LBT/MODS and 3.5m APO telescopes. The spectrum from the SDSS is also shown.
  Wavelengths are
in \AA\ and fluxes are in units of 10$^{-16}$ erg s$^{-1}$ cm$^{-2}$ \AA$^{-1}$.}
\label{fig7}
\end{figure*}
%%%%%%%%%%%%%%%%%%%%%%%%%%%%%%%%%%%%%%%%%%%%%%%%%

%%%%%%%%%%%%%%%%%%%%%%%%%%%%%%%%%%%%%%%%%%%%%%%%
%    Fig.8 timevar, 1420
%%%%%%%%%%%%%%%%%%%%%%%%%%%%%%%%%%%%%%%%%%%%%%%%
\begin{figure}
\hspace{0.1cm}\includegraphics[angle=-90,width=0.8\linewidth]{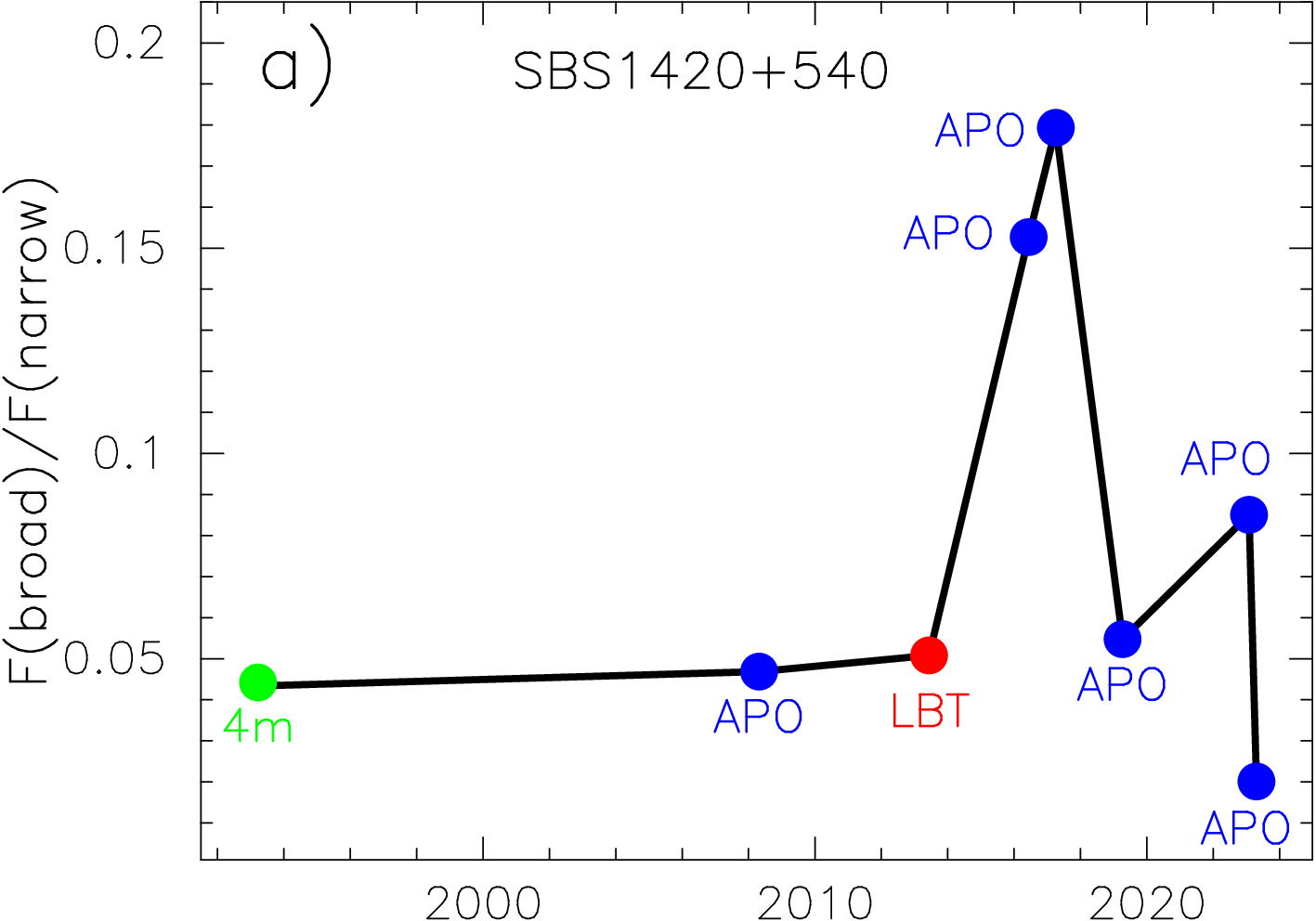}
\hspace{0.3cm}\includegraphics[angle=-90,width=0.8\linewidth]{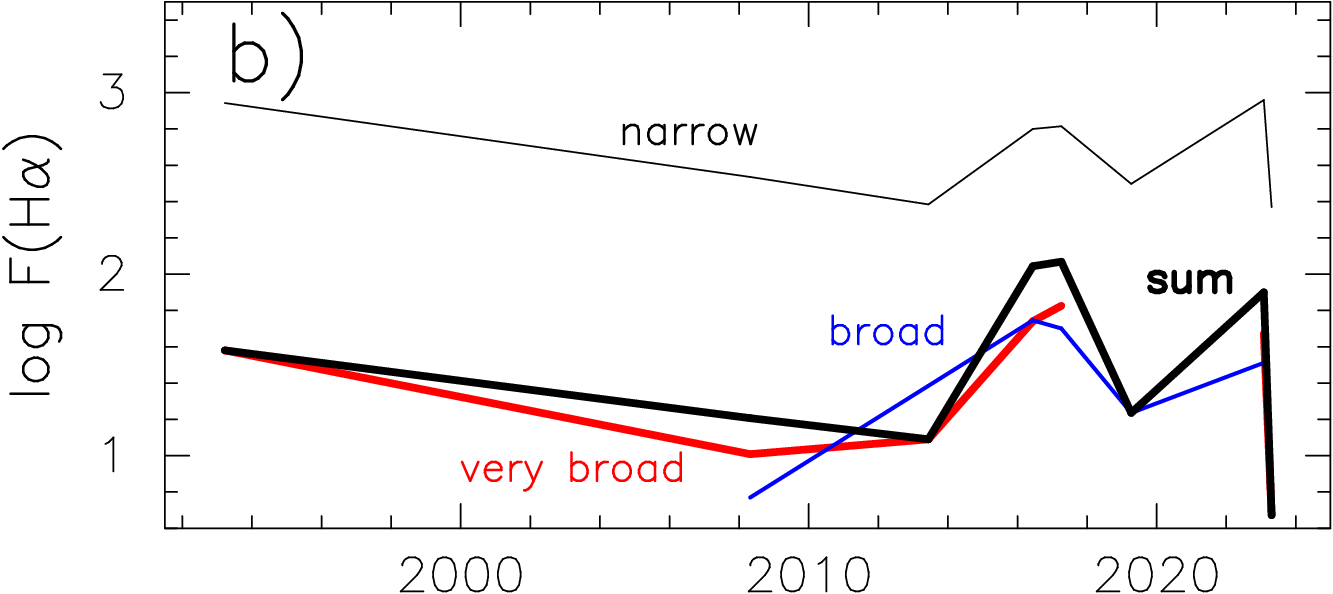} 
\hspace{0.3cm}\includegraphics[angle=-90,width=0.95\linewidth]{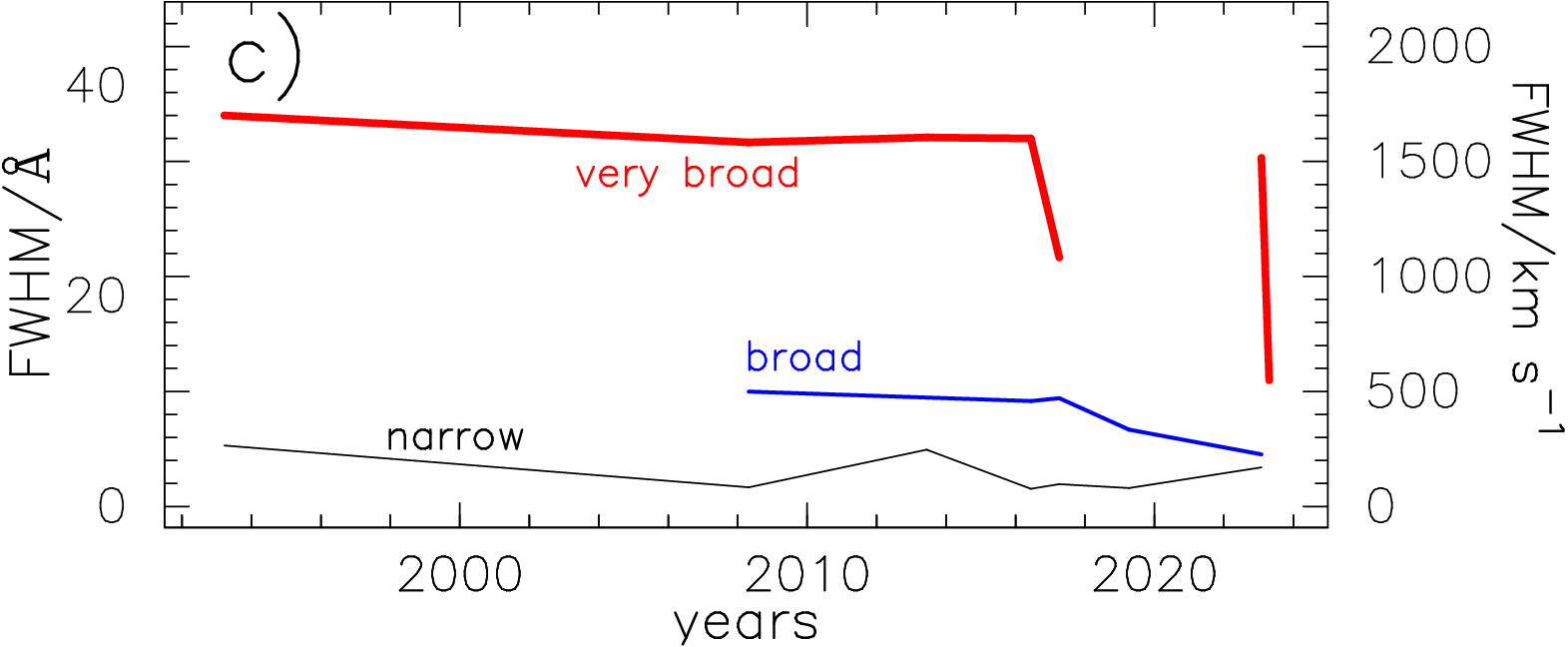}
\caption{{\bf a)} Temporal variations of H$\alpha$ broad-to-narrow flux ratio
 in SBS 1420$+$540.  
  Note that in {\bf a)} the broad component means the sum of the broad and very
broad emission, i.e. the whole broad bump.  
   {\bf b)} temporal variation of H$\alpha$ flux in units of log $F$(H$\alpha$) + 16, where $F$(H$\alpha$) is in erg s$^{-1}$cm$^{-2}$, of different components, narrow, broad and very broad, marked by different colours.
 The break in the sequence for the very broad component (red curve) is explained by the disappearance of the most fast-moving gas on 2019-04-08 and its reappearance in 2023.
  {\bf c)} The same as in {\bf b)} but for full width at half maximum in \AA\ (left scale) and in km s$^{-1}$ (right scale).
}
\label{fig8}
\end{figure}
%%%%%%%%%%%%%%%%%%%%%%%%%%%%%%%%%%%%%%%%%%%%%%%%%

%%%%%%%%%%%%%%%%%%%%%%%%%%%%%%%%%%%%%%%%%%%%%%%%
%    Fig.9 timevar,  1444
%%%%%%%%%%%%%%%%%%%%%%%%%%%%%%%%%%%%%%%%%%%%%%%%
\begin{figure}
\hspace{0.1cm}\includegraphics[angle=-90,width=0.8\linewidth]{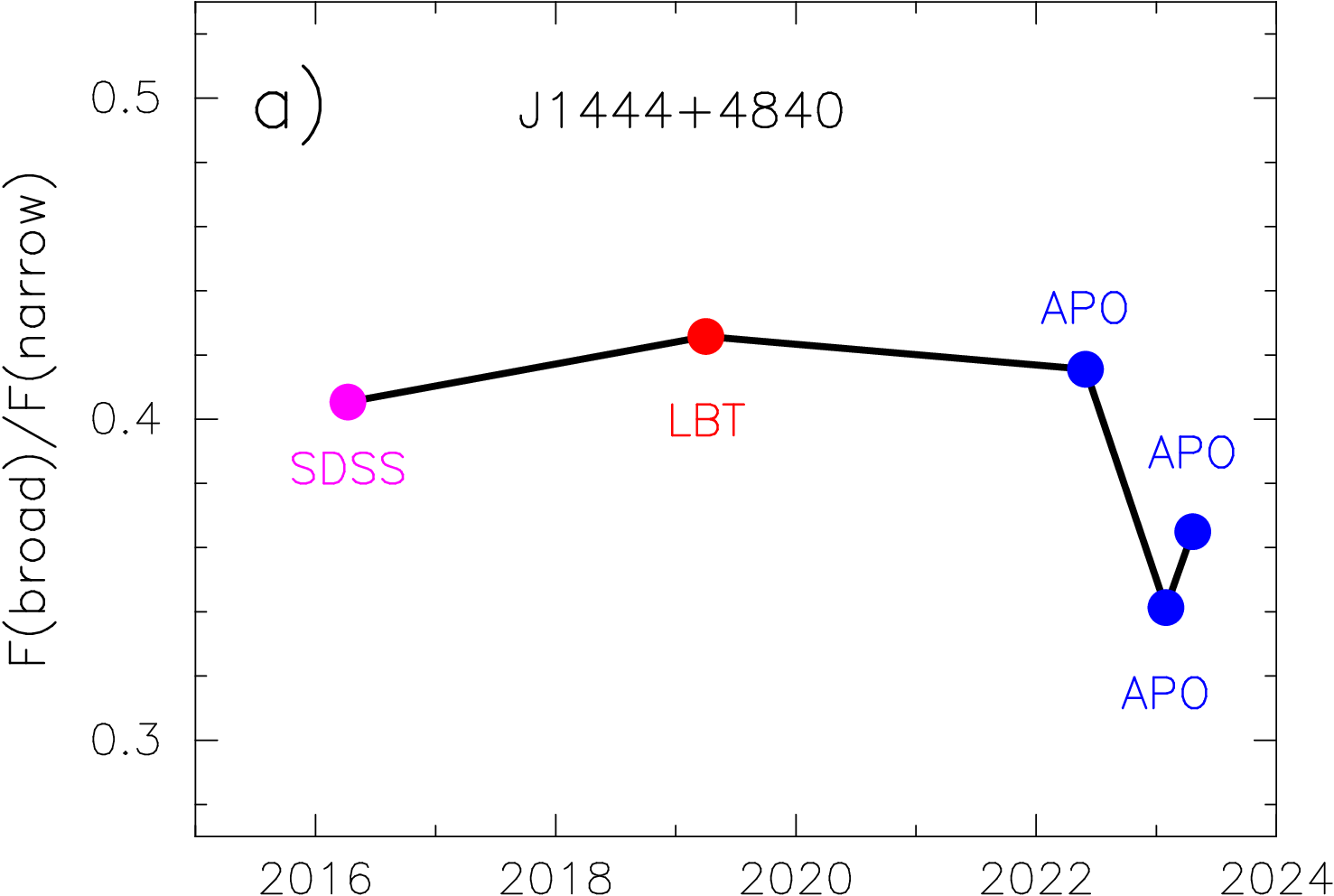}
\hspace{0.3cm}\includegraphics[angle=-90,width=0.8\linewidth]{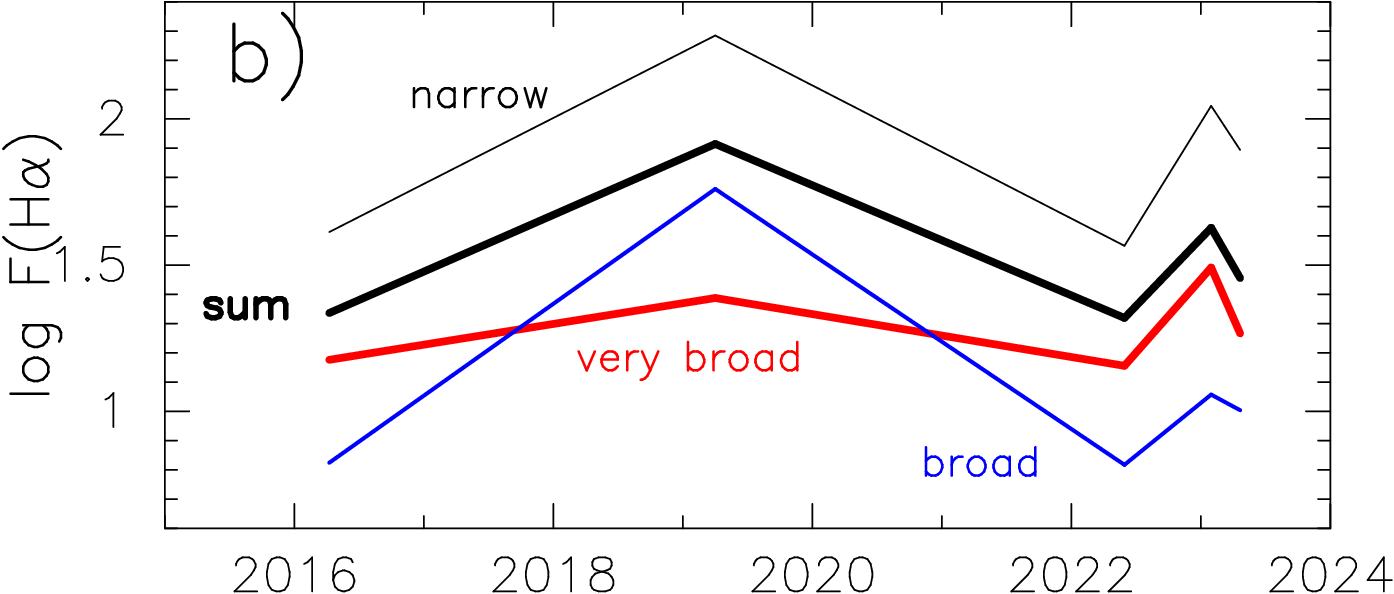}  
\hspace{0.3cm}\includegraphics[angle=-90,width=0.95\linewidth]{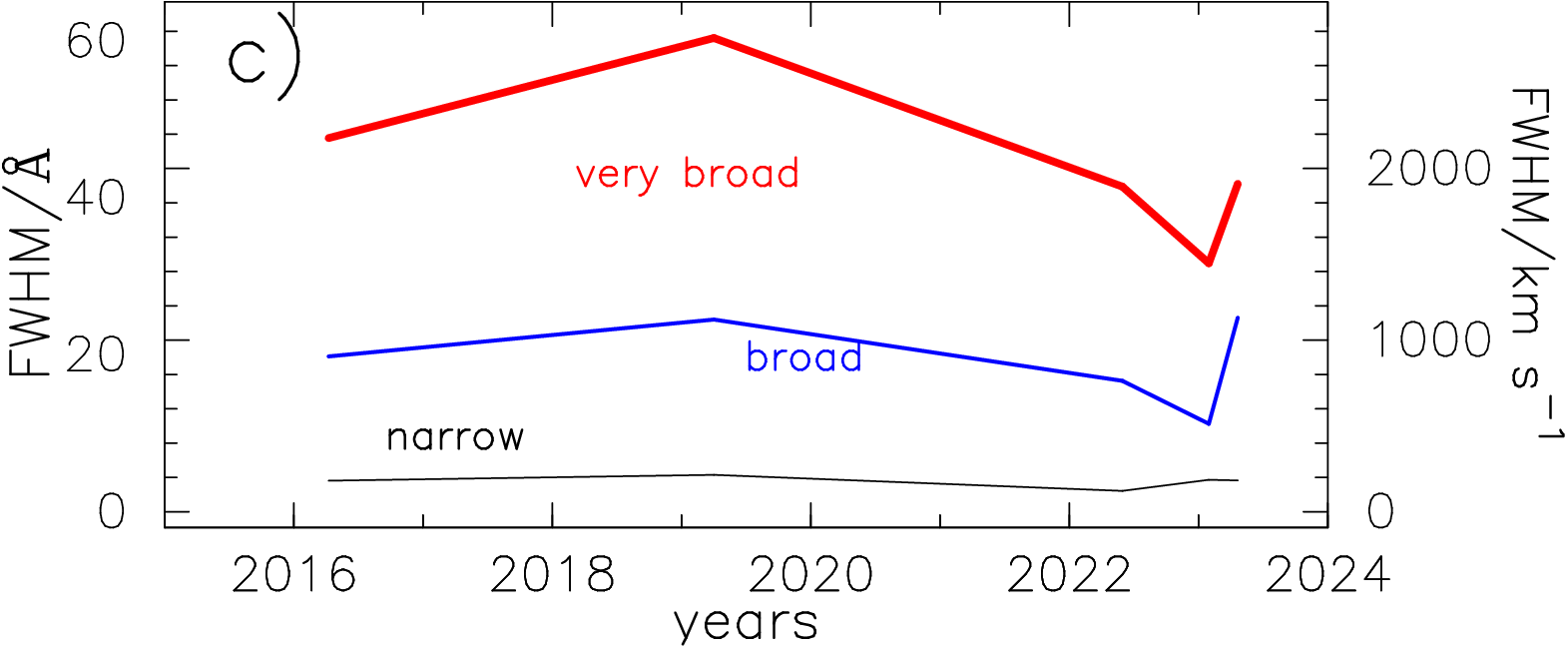}
\caption{The same as in Fig.~\ref{fig8} but for J1444$+$4840.
}
\label{fig9}
\end{figure}
%%%%%%%%%%%%%%%%%%%%%%%%%%%%%%%%%%%%%%%%%%%%%%%%%

%%%%%%%%%%%%%%%%%%%%%%%%%%%%%%%%%%%%%%%%%%%%%%%%%%%%%%%%%%%
%  Table 5
%%%%%%%%%%%%%%%%%%%%%%%%%%%%%%%%%%%%%%%%%%%%%%%%%%%%%%%%%%%
\begin{table*}
%\centering
\caption{Decomposition of strongest emission lines, LBT observations \label{tab5}}
\begin{tabular}{lcrrrrrrr} \hline
Line   & $\lambda$$^{\rm a}$ & \multicolumn{1}{c}{$F$$^{\rm b}$} & $L$$^{\rm c}$ & \multicolumn{1}{c}{$\Delta\lambda$$^{\rm d}$} &FWHM$^{\rm e}$ & $\sigma$$^{\rm f}$ & $v$$_{\rm term}$$^{\rm g}$ & Flux \\
       &  && \multicolumn{1}{c}{}&&&&&ratio$^{\rm h}$ \\
\hline
&\multicolumn{7}{c}{SBS 1420$+$544} \\
H$\beta$ (narrow)    &4861.4&   76.1&  7.8& 3.8&234 & 100 &  \\
H$\beta$ (absorb.)   &4843.3& $-$0.5&     & 8.6& & & \\
H$\beta$ (broad)     &4860.9&    3.5&  0.4&27.5&1697 & 721 & \\
H$\beta$ broad/narrow&        &                & &    & & &1090&0.046\\ \\
H$\alpha$ (narrow)   &6562.8&  242.2& 24.8& 5.4&247 & 105 & \\
H$\alpha$ (absorb.)  &6542.4& $-$2.4&     &20.4& & & \\
H$\alpha$ (broad)    &6560.4&   12.3&  1.3&35.1&1604 & 681 & \\
H$\alpha$ broad/narrow&       &                & &    & &  &823&0.050\\ \\
&\multicolumn{7}{c}{J1444$+$4840} \\
H$\beta$ (narrow)    &4860.2&   55.0& 56.3& 3.2&197 & 84 & \\
H$\beta$ (absorb.)   &4851.1& $-$0.6&     &12.5& & & \\
H$\beta$ (broad)     &4861.6&    7.7&  7.9&22.5&1389 & 590 &  \\
H$\beta$ broad/narrow&        &                & &   & &  &650&0.140\\ \\
He~{\sc i} (narrow)  &5874.9&    7.3&  7.5& 4.7&240 & 102&  \\
He~{\sc i} (absorb.) &5849.7& $-$0.3&     &10.2& & & \\
He~{\sc i} (broad)   &5875.8&    2.5&  2.6&32.4&1654 & 702& \\
He~{\sc i} broad/narrow&      &                & &   & &  &1330&0.338\\ \\
H$\alpha$ (narrow)   &6561.9&  192.6&197.0& 4.7&215 & 91 & \\
H$\alpha$ (absorb.)  &6525.0& $-$2.1&     &28.6& & & \\
H$\alpha$ (broad)    &6561.5&   57.6& 59.0&24.5&1120 & 476 & \\
H$\alpha$ (v.broad)  &6565.5&   24.4& 25.0&60.4&2761 & 1172 & \\
H$\alpha$ broad/narrow$^{\rm i}$&       &       & &    & &   &1670&0.426\\ \\
He~{\sc i} (narrow)  &7064.4&    4.4&  4.5& 4.8&204 & 86 & \\
He~{\sc i} (broad)   &7064.2&    3.0&  3.1&39.0&1656 & 703 & \\
He~{\sc i} broad/narrow&      &                & &       &   &  &&0.682\\ \\
\hline
  \end{tabular}

$^{\rm a}$Rest-frame wavelength in \AA.
$^{\rm b}$Observed flux in units 10$^{-16}$ erg s$^{-1}$cm$^{-2}$.
$^{\rm c}$Extinction-corrected luminosity in 10$^{39}$ erg s$^{-1}$.
$^{\rm d}$Full width at half maximum in \AA.
$^{\rm e}$Full width at half maximum in km s$^{-1}$.
$^{\rm f}$Velocity dispersion in km s$^{-1}$.
$^{\rm g}$The terminal velocity, $v$$_{\rm term}$ = $\lambda_{max}$(br) -- $\lambda_{min}$(abs) in km s$^{-1}$.
$^{\rm h}$Broad-to-narrow component flux ratio.
$^{\rm i}$Broad means the sum of the broad and very broad emission components.
  \end{table*}
%%%%%%%%%%%%%%%%%%%%%%%%%%%%%%%%%%%%%%%%%%%%%%%%%%%%%%%%%%%%%%%%%%%%

\section{Observations and data reduction}\label{sec:observations}

\subsection{LBT and APO observations}

  Coordinates, redshifts, and other characteristics of SBS\,1420$+$544 and J1444$+$4840 obtained from the SDSS photometric and spectroscopic databases are shown in Table \ref{tab1}.

We have obtained long-slit LBT spectra of
SBS\,1420$+$544 and J1444$+$4840 in 2013 and 2019, respectively. The observations were done 
in the twin binocular mode, using simultaneously the MODS1 and 
MODS2 spectrographs,
     equipped with CCDs (8022 pix $\times$ 3088 pix).
     The G400L blue grating (dispersion of 0.5\AA/pix), and the G670L red
grating (dispersion of 0.8\AA/pix) were used.
     The SBS 1420$+$544 spectrum covered the wavelength range
$\sim$3150 -- 9500\AA. With a 1.0 arcsec wide slit, 
this results in a resolving power $R$ $\sim$ 2000. 
    The seeing during the SBS 1420$+$544 observations was $\sim$ 1.0 arcsec. 
  Two subexposures were obtained for that galaxy, resulting in an effective
exposure time of 2$\times$1800s, after addition of MODS1 and MODS2 spectra.

  As for J1444$+$4840, the wavelength range is $\sim$3500 -- 8200\AA\ with
a 1.2 arcsec wide slit.
    The seeing during the J1444$+$4840 observations was also $\sim$ 1 arcsec.
       We obtained eight subexposures for this galaxy, resulting in a total exposure time of 2$\times$7200s.
    The logs of the observations are presented in Table \ref{tab2}.    

    Spectra of the spectrophotometric standard stars BD+28d4211 and Feige 34, 
 were also obtained during the night of observations 
with a 5 arcsec wide slit.
  They were used for flux calibration
and correction for telluric absorption lines. 
   We obtained also calibration frames of biases, flats and comparison lamps. 
Bias subtraction and flat field correction was done using 
the MODS Basic CCD Reduction package {\sc modsCCDRed} \citep{Pogge2019}.
   Other reductions, in particular flux and wavelength calibration were
performed with {\sc iraf}. 
Finally, all MODS1 and MODS2 subexposures were co-added, one-dimensional
spectra were extracted along the spatial axis by summing all the pixels with
emission in the aperture of 2.5 arcsec (Fig. \ref{fig2}a,b and Fig. \ref{fig3}).

In addition, a near-infrared (NIR) long-slit LBT spectrum of SBS 1420$+$544 was obtained in 2016 
in binocular mode, using LUCI1 and LUCI2. Our aim is 
to check whether the emission lines of hydrogen and helium in the near-infrared
range of 0.9--2.5$\mu$m covered by LUCI, 
have P-Cygni profiles.
     In addition, the NIR spectrum also allows the use of the density-sensitive infrared He {\sc i} $\lambda$10831\AA\ emission line for electron number density determination \citep{IzTG2014}. 

    The NIR observation was performed under good sky condition.
     We used the N1.8 camera with 0.25 arcsec pixel$^{-1}$ coupled with
the 60$\times$1.0 arcsec slit and the G200-LoRes grating, resulting in a spectral
resolving power of R~$\sim$~2000.
    Spectra have been reduced using {\sc iraf} and following the standard
long-slit reduction procedures described above.
    The near-infrared spectrum of SBS 1420$+$544 is shown on Fig. \ref{fig2}c in the wavelength range from 9400 to 13200\AA.

 In addition to the LBT observations, we also performed 
 supplementary observations with the 3.5m Apache Point Observatory (APO) telescope. 
The Dual Imaging Spectrograph (DIS) was used prior to 2023, 
and the Kitt Peak Ohio State Multi-Object Spectrograph (KOSMOS) in 2023.
  All APO observations 
were carried out in the long slit mode. For DIS we use 
 the red channel, with occasionally the blue channel.
Slit widths were in the range 0.9 -- 2.0 arcsec for DIS and equal to 1.18 arcsec for
KOSMOS. The average value of the airmass was 1.2.
   The DIS B1200 grating with a central wavelength of $\sim$6600 \AA\ and a linear dispersion of 0.56 \AA\ pixel$^{-1}$ was used in the red range, 
  providing a medium resolving power $R$ = 5000.
 {\sc iraf} routines were used for background subtractions, bias- and flat-field corrections.
    One-dimensional spectra were extracted after removal of cosmic particle trails and wavelength and flux calibration.
   All 
spectra used to study the spectral temporal variability are shown in Figs.~\ref{fig6} and ~\ref{fig7}.
  
 \subsection{Extinction-corrected emission-line fluxes} 

  The most striking feature of all spectra in Figs. \ref{fig2} and \ref{fig3} is the presence of many strong narrow emission lines. Among them, the hydrogen and helium lines stand out, the brightest of which showing also broad emission components, with the presence of blueward absorption, i.e. they display P-Cygni profiles.
  The P-Cygni profile is seen in the He {\sc i} $\lambda$5876 line. However, it is absent in the He {\sc i} lines $\lambda$4471 in the optical range and $\lambda$10831 in the near-infrared range.
    Note, however, that the NIR observation was carried out three years later than the optical ones and the absence of the P-Cygni profile may be due to flux variabilty. 
The [O~{\sc iii}]\,$\lambda$4363  emission line is clearly seen in these spectra.
  It allows to reliably determine the electron temperatures and the chemical composition of the two galaxies by the direct method.

  Emission-line fluxes and their errors were measured with the {\sc iraf} {\it splot} routine.
  For the determination of extinction from the Balmer decrements, we use the H$\alpha$, H$\beta$, H$\gamma$, H$\delta$, H9 -- H12 emission lines.
  Note that only the narrow components of hydrogen emission lines were measured in the cases when these lines also contain broad components.
    This means that the derived chemical composition is that for the ionized gas of the entire galaxy, excluding expanding circumstellar envelopes.
  
  All hydrogen line fluxes were corrected for extinction and underlying hydrogen 
stellar absorption, derived simultaneously in an iterative procedure, using the method described by \citet{ITL94}.
The equivalent widths of the underlying stellar Balmer absorption lines, EW(abs), in our iterative method are assumed to be the same for all hydrogen lines.
   The other lines  were corrected only for extinction.
  The ratios of the extinction-corrected nebular emission line fluxes to that of the narrow H$\beta$ line, together with the extinction coefficient $C$(H$\beta$), the observed H$\beta$ emission-line flux $F$(H$\beta$), the rest-frame equivalent width EW(H$\beta$), and the equivalent width of the Balmer absorption lines EW(abs) are shown for both galaxies in Table~\ref{tab3}.

  It is seen in the table that the extinction-corrected relative fluxes of the hydrogen emission lines are close to the theoretical relative recombination line intensities \citep[Table 8, p.86 in][]{HummerStorey1987}.
  Thus, we can be confident that the extinction coefficient $C$(H$\beta$) and  the equivalent width of the Balmer absorption lines EW(abs) were correctly calculated.

\section{Results}
\label{sec:results}   
   
\subsection{General properties} 
\label{sec:general}

The two studied galaxies are both dwarf SFGs, based on their low absolute magnitudes $M_g$ $\sim$ --17 mag, derived from the SDSS $g$ magnitude corrected for extinction in the Milky Way, and their low stellar masses $M_\star$ $\sim$ 2$\times$10$^6$ and 4$\times$10$^6$\,$M_{\odot}$ (Table~\ref{tab1}), derived from SED fitting of the aperture- and extinction-corrected SDSS spectra. They have 
low total 
extinction $A_V$ $\sim$ 0.4 and 0.6 mag for SBS 1420$+$540 and J1444$+$4840, respectively (Table~\ref{tab3}).
  Strong emission lines in the spectra, very high equivalent widths of hydrogen emission lines  [EW(H$\beta$) is equal to 328\AA\ and 288\AA\ for the two galaxies, as seen in Table~\ref{tab3}] suggest active star formation with the presence of a very young stellar population, with an age of $\sim$ 3 Myr.
  This is also evidenced by high star-formation rates and sSFRs (Table~\ref{tab1}) derived from extinction- and aperture-corrected H$\alpha$ luminosities, obtained from LBT spectra, and using equation 2 in \citet{K98}.

\subsection{Emission-line diagnostic diagrams}
\label{sec:diagnostic}

Strong emission lines are often used  to construct diagrams to diagnose different photoionization mechanisms in galaxies \citep{Steidel2014,Cullen2014,Topping2020,Backhaus2022}.
  One of the most used is the \citet*{BPT81} (BPT or O3N2) diagram where [O~{\sc iii}]$\lambda$5007/H$\beta$ is plotted against [N~{\sc ii}]$\lambda$6584/H$\alpha$. It  
separates galaxies into different classes (SFGs, AGNs or low-ionization narrow emission-line regions (LINERs)) according to their main excitation mechanisms.
  In the BPT diagram, low-$z$ SFGs define a tight relation called ``main sequence''. 
Such diagrams used to estimate the metallicities, ionization states, galaxies evolution with redshift and others.

  The locations of SBS~1420$+$544 and J1444$+$4840 in the BPT diagram
are indicated in Fig.~\ref{fig1}a.
   For comparison, compact SFGs from the SDSS 
  \citep{I16c} are also represented. 
  The two blue lines show the relations calculated with the 
{\sc cloudy} code c13.04 \citep{F98,F13} for metallicities 
12 $+$ log O/H = 7.3 and 8.0, starburst ages varying from 0 to 6 Myr and other input parameters 
 typical of low-metallicity SFGs \citep[see for details ][]{Iz2018}.

  Our two galaxies are located to the extreme left  
of the SFG main sequence, mainly because of their very low metallicities.
 This region to the left is also characteristic of objects with high ionization parameters
  \citep[see e.g. ][]{Steidel2014},  i.e. with young starburst ages.  
  For comparison, two previously studied galaxies with LBV candidates, PHL 293B and DDO 68 \citep{Pustilnik2008,IzT09,Guseva2022}, have also been plotted in
    Fig.~\ref{fig1}.
  However, note that the LBV in DDO 68 is located in the  H~{\sc ii} region number 3 (hereinafter \#3).
    The position of the most outlying  point (filled circle in the figure), DDO 68 \#3, is determined by its extremely low metallicity 12 $+$ log O/H $\sim$ 7.1 and its low equivalent width of H$\beta$, less than 100\AA\ \citep{IzT09}. The position of PHL 293B, with a metallicity 12 $+$ log O/H $\sim$ 7.7, is close to those 
of the two galaxies discussed here, but with its relatively low 
EW(H$\beta$) = 37 \AA.   

   Another type of diagram is the O$_{32}$ -- R$_{23}$ one (Fig.~\ref{fig1}b), where O32 = [O~{\sc iii}]5007/[O~{\sc ii}]3727 and R$_{23}$ = ([O~{\sc ii}]3727 + [O~{\sc iii}]4959 + [O~{\sc iii}]5007)/H$\beta$. Our two galaxies are similarly shifted from the main-sequence of compact SDSS SFGs to regions of lower metallicity, young age and high O$_{32}$, the latter  being an observational indicator of the ionization parameter.
For comparison, we have also shown in Fig.~\ref{fig1}b the locations of 
PHL 293B and H~{\sc ii} region \#3 in DDO 68 \citep{Guseva2022}. As for SBS 1420$+$540 and J1444$+$4840, their locations are shifted in both diagrams (Figs.~\ref{fig1}a and b). 
DDO 68 \#3 has the largest offset due to its extremely low metallicity.

\subsection{Element abundances}\label{subsec:abundances}

   The large wavelength range of our LBT spectra, from $\sim$3200\AA\ to $\sim$9500\AA\ for MODS, and from $\sim$3200\AA\ to $\sim$13000\AA\ for MODS and LUCI,  gives numerous bright emission lines. This allows us to derive abundances for many heavy elements.

   The presence of fairly intense oxygen emission lines [O~{\sc iii}] $\lambda$4363 in both objects makes it possible to use the {\it direct} $T_{\rm e}$ method for determining the metallicity of galaxies (expressed in units of 12 $+$ logO/H) with a high precision.
   We calculate the physical conditions and element abundances following the method described by \citet{IzStas2006}, 
based on the photoionization models of \citet{StasIz2003}.

 Briefly,  the electron temperature $T_{\rm e}$(O~{\sc iii}) is calculated from the ratio of oxygen emission line fluxes [O~{\sc iii}] $\lambda$4363/($\lambda$4959 + $\lambda$5007). 
  The electron temperatures $T_{\rm e}$(O~{\sc ii}) and $T_{\rm e}$(S~{\sc iii}) are then obtained from equations of \citet{IzStas2006}:
   \begin{equation}
 t({\rm O}~\textsc{ii})  = -0.744~+~t \times (2.338~-~0.610t),  \label{eq1}
   \end{equation}           

   \begin{equation}
    t({\rm S}~\textsc{iii}) = -1.276~+~t \times (2.645~-~0.546t),  \label{eq2}
      \end{equation}          
            where $t$(O~{\sc ii}) = 10$^{-4}$ $T_e$(O~{\sc ii}), $t$(S~{\sc iii}) = 10$^{-4}$ $T_e$(S~{\sc iii}) and $t$ = 10$^{-4}$ $T_e$(O~{\sc iii}).
 The electron number densities $N_{\rm e}$(S~{\sc ii}) are derived from the singly ionized sulfur line ratio [S~{\sc ii}] $\lambda$6717/$\lambda$6731. 
   We adopt $T_{\rm e}$(O~{\sc iii}) for O$^{2+}$, Ne$^{2+}$ and Ar$^{3+}$ abundance calculations. 
  $T_{\rm e}$(O~{\sc ii}) is used for O$^{+}$, N$^{+}$, S$^{+}$ and Fe$^{2+}$ abundance determination.
  $T_{\rm e}$(S~{\sc iii}) is adopted for the calculation of S$^{2+}$, Cl$^{2+}$ and Ar$^{2+}$ abundances.
   Since the high ionization line He~{\sc ii} $\lambda$4686\AA\ is present in the spectra, we can take into account unobservable O$^{3+}$:
   \begin{equation}
\frac{\rm O^{3+}}{\rm H^+} = 0.5 \times \frac{\rm He^{2+}}{\rm H^+ + \rm He^{2+}}~~(\frac{\rm O^{+}}{\rm H^+} + \frac{\rm O^{2+}}{\rm H^+}) \label{eq3}
   \end{equation}   
   To obtain the total oxygen abundance, we sum the abundances of the ions O$^{+}$, O$^{2+}$ and O$^{3+}$. 
  To take into account unobservable ionization states of other ions of heavy elements, we use the ionisation correction factors $ICF$s to obtain their total element abundances.

  The physical conditions in the H~{\sc ii} regions of the galaxies, i.e. the electron temperatures  $T_{\rm e}$ in the high-ionization zone O$^{2+}$, in the low-ionization zone O$^{+}$, and in the intermediate zone S$^{2+}$, together with their electron number densities $N_{\rm e}$(S~{\sc ii}), are given in Table~\ref{tab4}.
  We provide also in this table the ionic and total heavy element abundances for oxygen, nitrogen, neon, sulfur, chlorine, argon and iron, as well as their corresponding ionisation correction factors. 

  We find that both SBS 1420$+$544 and J1444$+$4840 have low metallicities, although not extremely low ones. They have respectively 12 + log O/H of 7.75 and 7.45, i.e. metallicities 
$Z$ $\sim$ 1/9 and 1/17 Z$_\odot$, if the solar oxygen abundance  12 + log O/H = 8.69  is adopted \citep{Asplund2021}.  
The ratios of other element abundances to oxygen abundance, specifically N/O, Ne/O, S/O, Cl/O, Ar/O and Fe/O, are similar to those usually derived for other low-metallicity star-forming galaxies \citep[see e.g. ][]{IzStas2006,Guseva2011,Guseva2020,Jeong2020,Senchyna2022}. 

  We do find an anomaly for J1444$+$4840:  
its N/O ratio is enhanced by $\sim$0.4 dex, compared to the lowest value $\sim$ $-$1.7 -- $-$1.6 for this metallicity.
   Can we understand this N/O enhancement? 
In the metallicity range 12 + logO/H $\la$ 8, the production of nitrogen is mainly primary so that 
the N/O ratio remains fairly constant, at a plateau value of about --1.5. 
        However, more and more SFGs with an enhanced N/O value have been found at the low metallicity end, both for low-$z$ \citep[see e.g. ][]{Amorin2012,Sanders2016,Vincenzo2016,Guseva2020}
        and high-$z$ \citep[see e.g. ][]{Kojima2017} galaxies. Several mechanisms have been proposed to account for this enhancement.
        \citet{IzStas2006}
proposed to explain the observed N/O
enrichment by a local N/O enhancement in dense
clumps of H~{\sc ii} regions,
created by winds of evolved WR stars.
        It has also been proposed that the N/O ratio increases as a result of the inflow of low-metalicity intergalactic gas onto the galaxy \citep{Amorin2010,Amorin2012,Loaiza2020}.
        Another proposal is the enhancement of nitrogen in rotating massive stars,  
as found in numerical simulations 
by \citet{Roy2020}, \citet{Grasha2021}, \citet{Roy2021}.
 \citet{Roy2022} also explain N/O enrichment in galaxies at high-redshifts using models with pre-supernova wind yields.
Lastly, both observations and models also suggest an enhanced nitrogen abundance due to the presence of LBVs in galaxies  \citep[see e.g. ][]{Weis2020}.

\subsection{Broad emission}
\label{sec:broad}

The most noticeable manifestations of a LBV eruption in the integrated spectrum of a star-forming dwarf galaxy are the observed strong broad emission components of hydrogen and sometimes of helium and metal lines.  The spectrum also often shows P-Cygni profiles.
 These broad, high-velocity components, superimposed on the narrow emission lines, are thought to arise in the circumstellar envelope and/or in a gas-dust nebula around a LBV star \citep[see e.g. ][]{Weis2020}.
  The narrow components are formed predominantly in the ionized interstellar medium of a galaxy and they very weakly increase in brightness 
during the eruptive LBV phase \citep[see e.g., the temporal variations of the narrow H$\alpha$ flux,  as compared to the broad one for DDO 68 \#3, shown by thin and thick black dotted lines in Fig. 7b by ][]{Guseva2022}.
  We can thus follow the temporal spectral variations caused by a LBV outburst by using as a tracer the flux ratio of the broad-to-narrow components of the brightest hydrogen and helium lines.
  Using that ratio allows us to bypass differences in exposure time, aperture and seeing when comparing observations taken with different telescopes in the course of the spectral monitoring (Figs.~\ref{fig6} and \ref{fig7}).

  The emission-line luminosities in Table \ref{tab5} have been derived with luminosity distances obtained from the observed redshifts (Table \ref{tab1}), using the NED cosmological calculator \citep{W06}.
  We have adopted the cosmological parameters $H_0$ = 67.1 km s$^{-1}$ Mpc$^{-1}$,
$\Omega_m$  = 0.318, $\Omega_{\Lambda}$ = 0.682 \citep{Planck2014}.

  We note that the luminosity $L$(H$\beta$) given in Table~\ref{tab5} for J1444$+$4840, is nearly 2 times lower than the value derived from the SDSS spectrum, both in the DR14 and DR16 releases. 
  A closer examination showed that the standard star for J1444$+$4840 was observed during non-photometric conditions, which can result in smaller calibrated fluxes for the galaxy.
  However, it should be remembered that the SDSS spectrum was obtained in 2016, while the LBT spectrum was taken in 2019, so the flux could have varied over this period.

  We have performed a decomposition of the strongest emission lines by fitting their profiles with multiple Gaussians (narrow, broad emission and absorption and also very broad emission, when needed), using the deblending routine of the {\sc iraf}/{\it splot} software package.
 Other authors  \citep[see e.g. in ][]{Iz2011,Terlevich2014} have also fitted high S/N H$\alpha$  profiles with more than two components.
   The results of the decomposition of the brightest hydrogen H$\alpha$ and H$\beta$ line profiles  in the LBT spectra of the two galaxies 
   are presented in Figs.~\ref{fig4} and \ref{fig5}.
   For J1444$+$4840, while the decomposition of the bright hydrogen lines 
can be made by using only a two-component fit, one narrow Gaussian and one broad Gaussian, the best fit to the high signal-to-noise profile of H$\alpha$ is
 given by a four-component fit.
In addition,  for J1444$+$4840, decomposition into two or three Gaussians has also been performed for the He~{\sc i} 5876 and 7065\AA\ lines. 
For the weaker lines of hydrogen and helium, only narrow and possible broad components can be distinguished.

  The results of the profile decompositions in the LBT/MODS spectra are also presented in Table~\ref{tab5}. Here, we give the observed fluxes, extinction-corrected luminosities and FWHMs for the two strongest hydrogen emission lines in both galaxies, and for the helium lines He~{\sc i} 5876 and 7065 \AA\ in J1444$+$4840.

\subsection{Signatures of LBVs}
\label{sec:signs}

   The properties of the two galaxies suggest that they contain a large number of massive stars, 
  the most massive of which can rapidly evolve to become LBV stars.
As remarked before, intense quasi-periodic variations in spectral features and  photometric brightness, the so-called S Dor variability or S Dor cycle which can last for years or decades, are the clearest  characteristic features that distinguish LBVs from other massive evolving stars \citep[see e.g. ][]{vanGenderen1997,Weis2020}.  In the spectra, these transient LBV phenomena manifest themselves by broad emission features with blueward absorption  in hydrogen and often in helium and Fe~{\sc ii} lines. 
  Our aim then is to search for time variations of these broad features during the monitoring period of the studied galaxies. 
We will discuss them in turm.

   \subsubsection{SBS 1420$+$540}
\label{sec:S1420}
   
   Permitted Fe~{\sc ii} emission, often observed in dense circumstellar envelopes of LBV stars, is 
not present in the LBT spectrum of SBS 1420$+$540.
   Instead, forbidden iron [Fe~{\sc iii}]\,$\lambda$$\lambda$4658, 4755, 4988, 5270\AA\ and  [Fe~{\sc v}] $\lambda$4227\AA\ emission lines 
are detected.

  As for the blueward absorption in P-Cygni profiles  near H$\alpha$,  it is not seen in every spectrum, but only in those with the highest S/N
(Figs.~\ref{fig4} and \ref{fig5}).  Furthermore, for 
 this particular galaxy, the absorption by the extended envelope occurs at wavelength $\sim$  6545\AA, close to the [N~{\sc ii}] $\lambda$6548\AA\ emission, 
  making it difficult to measure both with accuracy.

   Fig.~\ref{fig8}a shows the time variation of the broad-to-narrow line flux ratio of  H$\alpha$.
The flux ratio is seen to be nearly constant at 0.05 from 1993 to 2013, then to sharply increase by a factor of $\sim$ 4 after 2013, 
  reaching a maximum of $\sim$ 0.2 during the period 2016 -- 2017, before decreasing back to a level of about 0.02. This behavior means that we have caught SBS 1420$+$540
in a LBVc outburst phase, from beginning to end.

  It is seen from Fig.~\ref{fig8}b which displays  the temporal variations of the fluxes of the narrow, broad and very broad components of the H$\alpha$ emission-line, and their sum,  that   
  the outburst is characterized by an increase in  
  luminosity of the whole broad bump, i.e. of the sum of broad and very broad components.
The luminosity jumped by about one order of magnitude, from 1.3 to 12.0 $\times$ 10$^{39}$ erg s$^{-1}$,  typical of LBV luminosities.

   We have also examined the temporal variations of the kinematics of the ionized gas by measuring 
the full width at half-maximum (FWHM) of each Gaussian. The gas  
velocity dispersion $\sigma$ is then obtained by dividing the FHWM by 2.355 (Fig.~\ref{fig8}c).
   The FWHM measurements show 
that the velocity dispersion of the fastest-moving gas (the very broad component) reaches the  high value $\sigma$ $\sim$ 700 km s$^{-1}$ (or a FWHM  $\sim$ 1600 km s$^{-1}$) (Fig.~\ref{fig8}c).
It is interesting that the velocity dispersion of the fastest moving ionized matter remains fairly constant with time, during the period 1993--2019.
However, on 2019-04-08, the high-velocity emitting gas vanished abruptly after the outburst, its intensity becoming so low that it could not be detected anymore by our observations (Fig.~\ref{fig6}).
     Let us remember that after the burst of activity peaks in 2017, the broad-to-narrow flux ratio  also became very low.

   The latest observations in April 2023 show that the flux intensity (Fig.~\ref{fig8}b) and the velocity dispersion (Fig.~\ref{fig8}c) of the whole broad bump has sharply declined, after some fluctuation (Fig.~\ref{fig6}). As a result,  the broad-to-narrow flux ratio has decreased by an order of magnitude, compared to its maximum value  (Fig.~\ref{fig8}a).
 A similar picture of a sharp decrease of the velocity dispersion of the moving gas, shortly after the peak of activity, was also observed for the LBVc in DDO 68 \#3,  with a decrease of $\sigma$ from 500\,km\,s$^{-1}$ to 140\,km\,s$^{-1}$, and then to a complete disappearance of the broad component \citep{Guseva2022}.
  
    We calculate the stellar wind terminal velocity, $v$$_{\rm term}$. It  
is related to  $\Delta$$\lambda$ = $\lambda_{max}$(br) -- $\lambda_{min}$(abs), the difference between the wavelength at maximum intensity of the broad component and that of the  absorption component of a spectral profile.
We found the stellar wind terminal velocity to be $\sim$1000 km s$^{-1}$ for SBS 1420$+$540.

 In summary, the LBVc eruption in SBS 1420$+$540, with a peak luminosity of $\sim$ 10$^{40}$ erg s$^{-1}$, lasted for about 6 years, from 2013 to $\sim$2019, after which the intensity of the broad component returned back to its low original value, of about two hundredths of the intensity of the narrow component.
  However, the broad emission bump in the strong hydrogen and helium lines lasted much longer, persisting over the whole three-decade monitoring and still not disappearing completely (Figs.~\ref{fig2} and \ref{fig6}). 
 Thus, our spectroscopic monitoring  of SBS 1420$+$540 indicates that the LBVc
 has probably passed through the maximum of its eruption activity and is now likely on its way to the minimum.
   With continuing monitoring observations of SBS 1420$+$540, we will be able to further constrain the scenario of LBV stars undergoing periodic or quasi-periodic luminosity variations in this galaxy.

\subsubsection{J1444$+$4840}
\label{sec:J1444}

  A strong broad component, with a flux of approximately 40 per cent that of the narrow component, is clearly detected in the spectra of the compact SFG J1444$+$4840, during the whole period of its monitoring from 2016 to 2023.
This bright broad component is present not only in hydrogen lines, but also in helium lines.
  This is clearly seen for the He~{\sc i} $\lambda$5876\AA\ and $\lambda$7065\AA\ emission lines in Fig.~\ref{fig3}.
   The broad-to-narrow flux ratios are even slightly higher in helium lines than in hydrogen lines.
   We note that while all the broad-to-narrow flux ratios in J1444$+$4840 are one order of magnitude higher than those in SBS 1420$+$540, 
the broad-to-narrow flux ratios of helium lines in the latter are also similar to those of hydrogen lines.
  
   The temporal variations of the broad-to-narrow flux ratio in H$\alpha$ is shown in Fig.~\ref{fig9}a. We see 
 very small variations of this ratio over the 6 years from 2016 to 2022, with some small rise of activity in 2019, followed by a slight decrease of about 20 percent in 2023.
 Further monitoring is needed to know whether this signals the beginning of a decline in activity of a possible LBV star in this galaxy.

 Over the same period, the H$\alpha$ flux of the narrow component increased by a factor of 5 in 2019 compared to that in 2016  and 
2022 
(Fig.~\ref{fig9}b), 
 going from 40\,$\times$\,10$^{-16}$\,erg\,s$^{-1}$\,cm$^{-2}$ 
to $\sim$193\,$\times$\,10$^{-16}$\,erg\,s$^{-1}$\,cm$^{-2}$. 
  The flux of the broad component increased in 2019 by even more, $\sim$8 times.
There is also an increase of the H$\alpha$ flux in the very broad component, compared to 2016 and 2022.
  The flux increase of all velocity components in 2019 cannot be explained by an aperture effect, given that the slit width during LBT observations was 1.2 arcsec, and subsequent APO observations were made with comparable slit width of 1.5 arcsec (2022) and 1.18 arcsec (2023).

From 2019 to 2023, the velocity dispersion of the fastest-moving gas (very broad component) decreased by a factor of $\sim$2, from $\sim$1200 km s$^{-1}$ to $\sim$600 km s$^{-1}$ (Fig.~\ref{fig9}c).
   A similar behavior is seen for the broad component.
   The FWHM  of the fastest-moving gas, corresponding to velocities of 1400 - 2800\,km\,s$^{-1}$, stayed very high throughout the entire monitoring of the galaxy.

The luminosities of the broad H$\alpha$ bumps are high,  (2\,--\,8)\,$\times$\,10$^{40}$ ergs s$^{-1}$, higher than those in SBS 1420$+$540. They are similar to the values observed in active galactic nuclei (AGN) and Type IIn Supernovae (SNe) \citep{Kokubo2021,Burke2021}.
  High average terminal velocities, $v$$_{\rm term}$ $\sim$ 1200 km s$^{-1}$,  are seen.

 We now discuss the types of physical phenomena that can account for the above properties of J1444$+$4840: are they due to LBVc stars or to AGN/SN IIn?
  We first discuss dust extinction. LBVs are often associated with emission
  gas-dust nebulae as shown by both observations \citep[see e.g. ][]{Kniazev2015,Kniazev2016} and hydrodynamical simulations \citep{vanMarle2011}.
  The extinction coefficients derived from the observed
 hydrogen Balmer line ratios are small for our galaxies. They are $C$(H$\beta$) $\sim$ 0.2 (or $E$($B-V$) $\sim$ 0.1) for both (Table~\ref{tab3}).
 However, these extinction coefficients were obtained only from the narrow emission component. The $C$(H$\beta$) derived from the broad H$\alpha$ and H$\beta$ emission are significantly
   higher, by a factor of $\sim$5 -- 6 times, in any case if broad components are used  or if the sum of the broad and very broad components is used  (see Table~\ref{tab5}).
  Note, however, that a high Balmer decrement of the broad component can be caused not only by a high dust content, but also by a high density and thus by a high collisional excitation 
which is much higher 
for the H$\alpha$ emission line.
  By contrast, $C$(H$\beta$) for the broad component of SBS 1420$+$540 is approximately the same as that for the narrow component.
 
  A significant number of LBVs exhibit the presence of enhanced nitrogen  
 \citep[see e.g. ][]{Walborn2000,Weis2020}. This appears to be the case for J1444$+$4840.
 The [N~{\sc ii}] $\lambda$6548\AA\ emission line near H$\alpha$ is very weak and was not measured in J1444$+$4840 due to low metallicity (12 $+$ logO/H = 7.45) and due to
 the bright and broad emission and blueward absorption, which muffles the [N~{\sc ii}] nebular line.
 Nevertheless  N/O ratio is enhanced by $\sim$0.4 dex, as mentioned above in subsect. \ref{subsec:abundances}.
  The increased dustiness of the high velocity component of circumstellar ionized gas together with the enhanced ratio of N/O  
in J1444$+$4840 
may argue for the presence of LBVs in this galaxy.
 
 Permitted Fe~{\sc ii}  emission lines can act also as indicators of LBV stars. However, these permitted lines are absent in the LBT spectrum of J1444$+$4840.
  Only forbidden [Fe~{\sc ii}]\,$\lambda$4287\AA\ and [Fe~{\sc v}]\,$\lambda$4227\AA\ emission lines have been found whereas [Fe~{\sc iii}] lines seen in SBS 1420$+$540 are not detected.

   In summary, the spectral properties discussed above for J1444$+$4840 
can not definitely rule out the presence of a LBV star in it, but neither can they confirm it.
 During the whole 8-year monitoring period of J1444+4840, we have not observed significant variations of the broad-to narrow flux ratio, characteristic of eruptive events in LBVs.
  If we nevertheless adopt the LBV hypothesis, we would have to assume that the period of its eruptive cycle is larger than 8 years.  Again, further monitoring observations are needed to confirm that hypothesis.

\section{Conclusions}
\label{sec:conclus}

  We report the discovery and the time-monitoring of broad components with P-Cygni profiles of the hydrogen and helium emission lines in two compact star-forming galaxies
 (SFG): SBS 1420$+$540, selected from the Second Byurakan Survey (SBS), and J1444$+$4840, selected from the Sloan Digital Sky Survey (SDSS) Data Release 14 (DR14).
The spectrophotometric data have been obtained with the  2\,$\times$\,8.4\,m
Large Binocular Telescope (LBT) with the spectrographs MODS1, MODS2, LUCI1 and LUCI2 and the 3.5 m Apache Point Observatory (APO) telescope with the spectrographs DIS and KOSMOS.
  We collected also our different observations of the two galaxies obtained in previous studies.
  The spectroscopic data cover a time baseline of three decades for SBS~1420$+$544 and nearly one decade for J1444$+$4840.
 Our main results are the following:

  1) The two SFGs, SBS 1420$+$540 and J1444$+$4840, have low oxygen abundances  
12 $+$ logO/H = 7.75 $\pm$ 0.02 and 12\,$+$\,logO/H = 7.45 $\pm$ 0.02, respectively, derived by the {\it direct} $T_{\rm e}$ method.
  The N/O, Ne/O, S/O, Cl/O, Ar/O and Fe/O ratios are similar to those usually derived for other low-metallicity SFGs.
The N/O ratio for J1444$+$4840 
with lower oxygen abundance 
is enhanced by $\sim$0.4 dex compared to its average value.
 for given metallicity.
  
2)  We decompose the profiles of the brightest hydrogen and helium lines into the sum of three (four) Gaussians representing the narrow, broad, (very broad) and absorption components. 
  We use the flux ratio of the broad-to-narrow components to 
trace the temporal variations of possible LBV outbursts. 

3) For SBS 1420$+$540, our spectroscopic monitoring has captured the eruption phase of a LBVc (c for candidate) that lasted for about 6 years. 
The broad-to-narrow flux ratio sharply increased by a factor of 4 in 2017 and decreased by about an order of magnitude in 2023,  with a peak luminosity of the broad component $L$(H$\alpha$) $\sim$ 10$^{40}$ erg s$^{-1}$.  
  The velocity dispersion of the fastest moving gas reaches the value $\sigma$ $\sim$ 700 km s$^{-1}$ (or FWHM $\sim$ 1600 km s$^{-1}$), and the terminal velocity of the stellar wind, derived from the P-Cygni features, is $v$$_{\rm term}$ $\sim$1000 km s$^{-1}$.
Our spectroscopic monitoring of SBS 1420+540 indicates that the LBVc has probably passed through the maximum of its eruption activity.

 4) For J1444$+$4840, the fluxes of the strong broad components of hydrogen and helium lines are quite high, amounting to $\sim$40 per cent that of the narrow components. 
 The H$\alpha$ broad component shows a high luminosity $L$(H$\alpha$) about 10$^{41}$ ergs s$^{-1}$, a value that is more similar to those observed in active galactic nuclei (AGNs) and Type IIn Supernovae (SNe) than in LBVs.
We found the variability of less than 20 per cent of the broad-to-narrow flux ratios, the high velocity dispersion $\sigma$ $\sim$600 -- 1200 km s$^{-1}$, and the high terminal velocity $v$$_{\rm term}$ $\sim$1000 -- 1700 km s$^{-1}$ that has persisted over 8 years of monitoring.
The results obtained thus far do not allow us to definitively conclude that a LBVc is present in J1444$+$4840.  The hypothesis of a long-lived stellar transient of type AGN/SN IIn would work as well.

\section*{Acknowledgements}

N.G.G. and Y.I.I. acknowledge support from
the National Academy of Sciences of Ukraine (Project No. 0121U109612
``Dynamics of particles and collective excitations in high energy physics,
astrophysics and quantum microsystem'') and from the Simons Foundation. T.X.T. thanks the hospitality of the 
Institut d'Astrophysique in Paris where part of this work was carried out.
The Large Binocular Telescope (LBT)
is an international collaboration among institutions in the United States, 
Italy and Germany. LBT Corporation partners are: The University of Arizona on 
behalf of the Arizona university system; Istituto Nazionale di Astrofisica, 
Italy; LBT Beteiligungsgesellschaft, Germany, representing the Max-Planck 
Society, the Astrophysical Institute Potsdam, and Heidelberg University; The
Ohio State University, and The Research Corporation, on behalf of The
University of Notre Dame, University of Minnesota and University of Virginia.
This paper used data obtained with the Multi-Object Double Spectrograph (MODS)
built with funding from NSF grant AST-9987045 and the NSF Telescope System
Instrumentation Program (TSIP), with additional funds from the Ohio
Board of Regents and the Ohio State University Office of Research.
{\it LBT} Near Infrared Spectroscopic {\it Utility} with {\it Cameras} and
{\it Integral} Field Unit for Extragalactic Research (LUCI1 and LUCI2) are
mounted at the front Bent Gregorian
f/15 focal stations of the LBT, and were designed and built by
the LUCI consortium, which consists of Landessternwarte Heidel-
berg, MPE, MPIA, and Astronomisches Institut der Universitt
Bochum.
Image Reduction and Analysis Facility ({\sc iraf}) is distributed by the 
National Optical Astronomy Observatories, which are operated by the Association
of Universities for Research in Astronomy, Inc., under cooperative agreement 
with the National Science Foundation.
Funding for the Sloan Digital Sky Survey IV has been provided by
the Alfred P. Sloan Foundation, the U.S. Department of Energy Office of
Science, and the Participating Institutions. SDSS-IV acknowledges
support and resources from the Center for High-Performance Computing at
the University of Utah. The SDSS web site is www.sdss.org.
SDSS-IV is managed by the Astrophysical Research Consortium for the 
Participating Institutions of the SDSS Collaboration. 
This research has made use of the NASA/IPAC Extragalactic Database (NED), which 
is operated by the Jet Propulsion Laboratory, California Institute of 
Technology, under contract with the National Aeronautics and Space 
Administration.This work was supported by a grant from the Simons Foundation (Grant Number 1030283, N.G.G. and Y.I.I.).

\section{DATA AVAILABILITY}

The data underlying this article will be shared on reasonable request to the corresponding author.

\bsp

\label{lastpage}


\begin{thebibliography}{}

\bibitem[Ahumada et al.(2020)]{Ahumada2020} Ahumada R., Allende Prieto C.,
  Almeida A., et al. 2020, \apjs, 249, 1

\bibitem[Amorin et al.(2010)]{Amorin2010}  Amorin R. O., P\'erez-Montero E., V\'ilchez J. M., 2010, \apj, 715, L128

\bibitem[Amorin et al.(2012)]{Amorin2012} Amorin R. O., V\'ilchez J. M.,
  P\'erez-Montero E., 2012, \assp, 28, 243

\bibitem[Annibali et al.(2019a)]{Annibali2019} Annibali F. et al., 2019a, \mnras, 482, 3892 

\bibitem[Annibali et al.(2019b)]{Annibali2019a} Annibali F. et al., 2019b, \apj, 883, 19   

\bibitem[Asplund et al.(2021)]{Asplund2021} Asplund M., Amarsi A. M.,
  \& Grevesse N., 2021, \aap, 653A, 141A

\bibitem[Baldwin et al.(1981)Baldwin, Phillips \& Terlevich]{BPT81} 
Baldwin J. A., Phillips M. M., Terlevich R., 1981, \pasp, 93, 5

\bibitem[Backhaus et al.(2022)]{Backhaus2022} Backhaus B. E., Trump J. R., Cleri N. J., Simons R., Momcheva I., Papovich, C. et al., 
2022,\apj, 926, 161B 

\bibitem[Bestenlehner et al.(2023)]{Bestenlehner2023} Bestenlehner J. M.,  Enßlin T., Bergemann M., Crowther P. A., Greiner M., Selig M., et al., 2023, arXiv:2309.06474

\bibitem[Burke et al.(2021)]{Burke2021} Burke C. J., Liu X., Chen Y.-C.,
  Shen Y., Guo H., 2021, \mnras, 504, 543 
  
\bibitem[Crowther \& Hadfield(2006)]{CrowtherHadfield2006} Crowther P. A.,
  Hadfield L. J., 2006, \aap, 449, 711

\bibitem[Crowther(2007)]{Crowther2007} Crowther P. A., 2007, \araa, 45, 177

\bibitem[Cullen et al.(2014)]{Cullen2014} Cullen F., Cirasuolo M., McLure R. J., Dunlop J. S., and Bowler R. A. A., 2014, \mnras, 440, 2300

\bibitem[Davidson(1999)]{Davidson1999} Davidson K., 1999, \aspc, 179, 6 

\bibitem[Davidson \& Humphreys(1997)]{DavidsonHumphreys1997}Davidson K.,
  Humphreys R. M., 1997, \araa, 35, 1

\bibitem[Drissen et al.(1997)Drissen, Roy \& Robert]{Drissen1997} Drissen L., Roy J.-R., Robert C.,
  1997, \apj, 474, L35

\bibitem[Drissen et al.(2001)]{Drissen2001} Drissen L., Crowther P. A.,
  Smith L. J., Robert C., Roy J.-R., Hillier D. J., 2001, \apj, 546, 484

\bibitem[Ferland et al.(1998)]{F98} Ferland G. J., Korista K. T.,
Verner D. A., Ferguson J. W., Kingdon J. B., Verner E. M., 1998,
\pasp, 110, 761

\bibitem[Ferland et al.(2013)]{F13} Ferland G. J. et al., 2013, \rmxaa, 
49, 137

\bibitem[Grasha et al.(2021)]{Grasha2021} Grasha K., Roy A., Sutherland R. S. \&  Kewley L. J., 2021, \apj, 908, 241

\bibitem[Grassitelli et al.(2020)]{Grassitelli2020} Grassitelli L, Langer N.,
  Mackey J., et al., 2020, arXiv:2012.00023

\bibitem[Guseva et al.(2011)]{Guseva2011} Guseva N. G., Izotov Y. I.,
  Stasi\'nska G., Fricke K. J., Henkel C., Papaderos P.,
  2011, \aap, 529, A149

\bibitem[Guseva et al.(2020)]{Guseva2020} Guseva N. G. et al., 
 2020, \mnras, 497, 4293     

\bibitem[Guseva et al.(2022) Guseva, Thuan \& Izotov]{Guseva2022}
Guseva N. G., Thuan T. X., Izotov Y. I. 2022, \mnras, 512, 4298 
  
\bibitem[Hummer \& Storey(1987)]{HummerStorey1987} Hummer D. G., Storey P. J., 1987, \mnras, 224, 801

\bibitem[Humphreys \& Davidson(1994)]{HumphreysDavidson1994} Humphreys R. M.,
 Davidson K. 1994, \pasp, 106, 1025 

\bibitem[Humphreys et al.(2013)]{Humphreys2013} Humphreys R. M., Davidson K.,
  Grammer S., et al., 2013, \apj, 773, 46

\bibitem[Humphreys et al.(2017)]{Humphreys2017} Humphreys R. M. et al., 2017, \apj, 844, 40

\bibitem[Humphreys (2019)]{Humphreys2019} Humphreys R. M., Introductory review
  chapter for e-book "Luminous stars in Nearby Galaxies"
  Galaxies, vol. 7, issue 3, p. 75, 2019; doi:10.3390/galaxies7030075

\bibitem[Izotov \& Thuan(2008)]{IzT08}
Izotov Y. I., Thuan T. X., 2008, \apj, 687, 133 

\bibitem[Izotov \& Thuan(2009)]{IzT09}
Izotov Y. I., Thuan T. X., 2009, \apj, 690, 1797 

\bibitem[Izotov et al.(1994)Izotov, Thuan \& Lipovetsky]{ITL94} Izotov Y. I.,
Thuan T. X., Lipovetsky V. A., 1994, \apj, 435, 647

\bibitem[Izotov et al.(2006)]{IzStas2006} Izotov Y. I., Stasi\'nska G.,
 Meynet G., Guseva N. G., Thuan T. X., 2006, \aap, 448, 955

\bibitem[Izotov et al.(2007)Izotov, Thuan \& Guseva]{IzTG07}
  Izotov Y. I., Thuan T. X., Guseva N. G., 2007, \apj, 671, 1297 

\bibitem[Izotov et al.(2011)]{Iz2011} Izotov Y. I., Guseva N. G., Fricke K. J., Henkel C., 2011, \aap, 533, A25

\bibitem[Izotov et al.(2014)Izotov, Thuan \& Guseva]{IzTG2014} Izotov Y. I., 
Thuan T. X., Guseva N. G., 2014, \mnras, 445, 778 

\bibitem[Izotov et al.(2016)]{I16c} Izotov Y. I., Guseva N. G.,
Fricke K. J., Henkel C., 2016, \mnras, 462, 4427 

\bibitem[Izotov et al.(2018a)]{Iz2018} Izotov Y. I., Thuan T. X., Guseva N. G.,  Liss S. E., 2018a, \mnras, 473, 1956  

\bibitem[Jeong et al.(2020)]{Jeong2020} Jeong M.-S., Shapley A. E., Sanders R. L., Runco J. N., Topping M. W., Reddy N. A., et al., 2020, \apj, 902, 16

\bibitem[Kauffmann et al.(2003)]{K03} Kauffmann G. et al., 2003, \mnras, 341, 33

\bibitem[Kennicutt(1998)]{K98} Kennicutt R. C., Jr.,
1998, \araa, 36, 189

\bibitem[Kniazev et al.(2015)]{Kniazev2015} Kniazev A. Y., Gvaramadze V. V., Berdnikov L. N., 2015, \mnras, 449, L60

\bibitem[Kniazev et al.(2016)]{Kniazev2016} Kniazev A. Y., Gvaramadze V. V., Berdnikov L. N., 2016, \mnras, 459, 3068

\bibitem[Kojima et al.(2017)]{Kojima2017} Kojima T., Ouchi M., Nakajima K., 
  Shibuya T., Harikane Y., Ono Y., 2017, \pasj, 69, 44

\bibitem[Kokubo (2021)]{Kokubo2021} Kokubo M., 2021, \mnras, in press;
  arXiv:2101.07797   

\bibitem[Lamers et al.(1983)]{Lamers1983} Lamers H. J. G. L. M., de Groot M.,
 Cassatella A., 1983, \aap, 128, 299 

\bibitem[Leitherer et al.(2001)]{Lei01} Leitherer C., Le\=ao J. R. S., 
Heckman T. M., Lennon D. J., Pettini M., Robert C., 2001, \apj, 550, 724

\bibitem[Loaiza-Agudelo, Overzier \& Heckman(2020)]{Loaiza2020} Loaiza-Agudelo M., Overzier R. A., Heckman T., 2020, \apj, 891, 19

\bibitem[Massey et al.(2000)]{Massey2000} Massey P., Waterhouse E.,
  DeGioia-Eastwood K., 2000, \aj, 119, 2214

\bibitem[Planck Collaboration XVI(2014)]{Planck2014} Planck Collaboration XVI,
2014, \aap, 571, A16

\bibitem [Pogge(2019)]{Pogge2019} Pogge R., 2019, rwpogge/modsCCDRed: v2.0.1,
Basic 2D CCD Reduction Programs for the Multi-Object Double Spectrographs
(MODS) at the Large Binocular Telescope Observatory, Available at:https://doi.org/10.5281/zenodo.2647501

\bibitem[Pustilnik et al.(2008)]{Pustilnik2008} Pustilnik S. A.,
  Tepliakova A. L., Kniazev A. Y., Burenkov A. N., 2008, \mnras, 388, L24 

\bibitem[Pustilnik et al.(2017)]{Pustilnik2017} Pustilnik S. A., Makarova,
L. N., Perepelitsyna, Y. A., Moiseev A. V., Makarov D. I.,
2017, \mnras, 465, 4985 

\bibitem[Roy et al.(2020)]{Roy2020} Roy A., Sutherland R. S., Krumholz M. R., Heger A. \& Dopita M. A., 2020, \mnras, 494, 3861

\bibitem[Roy et al.(2021)]{Roy2021} Roy A., Dopita M. A., Krumholz M. R.,  Kewley, L. J., Sutherland R. S., Heger A., 2021, \mnras, 502, 4359

\bibitem[Roy et al.(2022)]{Roy2022} Roy A., Krumholz M. R., Dopita M. A., Sutherland R. S., Kewley, L. J., Heger A., 2022, IAUS, 366, 33R %Proceedings of the International Astronomical Union, Volume 366, pp. 33-38

\bibitem[Sanders et al.(2016)]{Sanders2016} Sanders R. L., Shapley A. E.,
  Kriek M., Reddy N. A., Freeman W. R., Coil A. L., et al., 2016, \apj, 816, 23
  
\bibitem[Schaerer et al.(2000)]{SchaererGus2000}
Schaerer D., Guseva N. G., Izotov Y. I., Thuan T. X., 2000, \aap, 362, 53

\bibitem[Schaerer \& Vacca(1998)]{SchaererVacca1998} Schaerer D., Vacca W. D.,
  1998, \apj, 497, 618 

\bibitem[Senchyna et al.(2022)]{Senchyna2022} Senchyna P.,  Stark D. P., Charlot S., Plat A., Chevallard J.,  Chen Z., et al., 2022, \apj, 930, 105

\bibitem[Smith et al.(1994)]{Smith1994} Smith L. J., Crowther P. A.,
  Prinja R. K. 1994, \aap, 281, 833

\bibitem[Smith et al.(2011)]{Smith2011} Smith N., Li W., Silverman J. M.,
  Ganeshalingam M., Filippenko A. V., 2011, \mnras, 415, 773

\bibitem[Solovyeva(2020)]{Solovyeva2020} Solovyeva Y. et al., 2020,  arXiv:2008.06215

\bibitem[Stasi\'nska \& Izotov(2003)]{StasIz2003} Stasi\'nska G., Izotov, Y. I., 2003, \aap, 397, 71

\bibitem[Steidel et al.(2014)]{Steidel2014} Steidel C. C. et al., 2014, \apj, 795, 165

\bibitem[Terlevich et al.(2014)]{Terlevich2014} Terlevich R., Terlevich E., Bosch G., D\'iaz \'A., H\"agele G., Cardaci M., Firpo V., 2014, \mnras, 445, 1449

\bibitem[Topping et al.(2020)]{Topping2020} Topping M. W. et al., 2020, \mnras, 499, 1652T  

\bibitem[van Genderen et al.(1997)]{vanGenderen1997} van Genderen A. M.,
  Sterken  C., de Groot M., 1997, \aap, 318, 81

\bibitem[van Marle et al.(2011)]{vanMarle2011} van Marle A. J., Meliani Z.,
  Keppens R. Decin L., 2011, \apj, 734, L26, [arXiv:astro-ph.SR/1105.2387]
doi:10.1088/2041-8205/734/2/L26

\bibitem[Vincenzo et al.(2016)]{Vincenzo2016} Vincenzo F., Belfiore F.,
  Maiolino R., Matteucci F., Ventura P., 2016, \mnras, 458, 3466

\bibitem[Vink(2023)]{Vink2023} Vink J. S., Mehner A., Crowther P. A., Fullerton A., Garcia M., Martins F., et al., 2023, \aap, 675, 154

\bibitem[Vink(2012)]{Vink2012} Vink J. S., 2012, \assl, 384, 221 

\bibitem[Walborn \& Fitzpatrick(2000)]{Walborn2000} Walborn N. R., Fitzpatrick E. L., 2000, \pasp, 112, 50
doi:10.1086/316490

\bibitem[Weis \& Bomans(2020)]{Weis2020} Weis K., Bomans D. J., 2020, Review,
  arXiv:2009.03144

\bibitem[Wright(2006)]{W06} Wright E. L., 2006, \pasp, 118, 1711  
  

\end{thebibliography}
\end{document}